\documentclass[11pt,twocolumn]{aastex63}

\received{June 11, 2021}
\accepted{September 5, 2021}
\submitjournal{ApJ}

\newcommand{\pmra}{$\mu_\alpha \cos\delta$}
\newcommand{\pmdec}{$\mu_\delta$}

\shorttitle{Hierarchical clustering in Vela OB2}
\shortauthors{Pang et al.}

\begin{document}

\title{Disruption of Hierarchical Clustering in the Vela OB2 Complex and \\
                the Cluster Pair Collinder\,135 and UBC\,7 with Gaia EDR3: Evidence of Supernova Quenching}

\author[0000-0003-3389-2263]{Xiaoying Pang}
    \affiliation{Department of Physics, Xi'an Jiaotong-Liverpool University, 111 Ren’ai Road, 
                Dushu Lake Science and Education Innovation District, Suzhou 215123, Jiangsu Province, P. R. China, Xiaoying.Pang@xjtlu.edu.cn}
    \affiliation{Shanghai Key Laboratory for Astrophysics, Shanghai Normal University, 
                100 Guilin Road, Shanghai 200234, P. R. China}

\author[0000-0001-6980-2309]{Zeqiu Yu}
    \affiliation{Department of Physics, Xi'an Jiaotong-Liverpool University, 111 Ren’ai Road, 
                Dushu Lake Science and Education Innovation District, Suzhou 215123, Jiangsu Province, P. R. China, Xiaoying.Pang@xjtlu.edu.cn}

\author[0000-0003-4247-1401]{Shih-Yun Tang}
    \affiliation{Lowell Observatory, 1400 W. Mars Hill Road, Flagstaff, AZ 86001, USA}
    \affiliation{Department of Astronomy and Planetary Sciences, Northern Arizona University, Flagstaff, AZ 86011, USA}

\author[0000-0003-3784-5245]{Jongsuk Hong}
    \affiliation{Korea Astronomy and Space Science Institute, Daejeon 34055, Republic of Korea}

\author[0000-0002-8129-5415]{Zhen Yuan}
    \affiliation{Universit\'e de Strasbourg, CNRS, Observatoire Astronomique de Strasbourg, UMR 7550, F-67000 Strasbourg, France}

\author[0000-0003-3784-5245]{Mario Pasquato}
    \affiliation{Center for Astro, Particle and Planetary Physics (CAP$^3$), New York University Abu Dhabi}
    \affiliation{INFN- Sezione di Padova, Via Marzolo 8, I–35131 Padova, Italy}

\author[0000-0002-1805-0570]{M.B.N. Kouwenhoven}
    \affiliation{Department of Physics, Xi'an Jiaotong-Liverpool University, 111 Ren’ai Road, 
                Dushu Lake Science and Education Innovation District, Suzhou 215123, Jiangsu Province, P. R. China, Xiaoying.Pang@xjtlu.edu.cn}

\begin{abstract} 

We identify hierarchical structures in the Vela OB2 complex and the cluster pair Collinder\,135 and UBC\,7 with Gaia EDR3 using the neural network machine learning algorithm \texttt{StarGO}. Five second-level substructures are disentangled in Vela OB2, which are referred to as Huluwa\,1 (Gamma Velorum), Huluwa\,2, Huluwa\,3, Huluwa\,4 and Huluwa\,5. For the first time, Collinder\,135 and UBC\,7 are simultaneously identified as constituent clusters of the pair with minimal manual intervention. We propose an alternative scenario in which Huluwa\,1--5 have originated from sequential star formation. The older clusters Huluwa\,1--3 with an age of  10--22\,Myr, generated stellar feedback to cause turbulence that fostered the formation of the younger-generation Huluwa\,4--5  (7--20\,Myr). A supernova explosion located inside the Vela IRAS shell quenched star formation in Huluwa\,4--5 and rapidly expelled the remaining gas from the clusters. This resulted in global mass stratification across the shell, which is confirmed by the regression discontinuity method. The stellar mass in the lower rim of the shell is $0.32\pm0.14$\,$\rm M_\odot$ higher than in the  upper rim. Local, cluster-scale mass segregation is observed in the lowest-mass cluster Huluwa\,5. Huluwa\,1--5 (in Vela OB2) are experiencing significant expansion, while the cluster pair suffers from moderate expansion. The velocity dispersions suggest that all five groups (including Huluwa\,1A and Huluwa\,1B) in Vela OB2 and the cluster pair are supervirial and are undergoing disruption, and also that Huluwa\,1A and Huluwa\,1B may be a coeval young cluster pair.  $N$-body simulations predict that Huluwa\,1--5 in Vela OB2 and the cluster pair will continue to expand in the future 100\,Myr and eventually dissolve.

\end{abstract}

\keywords{stars: evolution --- open clusters and associations: individual -- stars: kinematics and dynamics -- methods: statistical -- methods: numerical }

\section{Introduction}\label{sec:intro}

Star clusters tend to form in complexes \citep{efremov1978}. A typical complex contains several OB associations and a few young clusters, with a star formation process lasting from a few to a dozen million years. 
Stellar complexes tend to be hierarchical, with smaller substructures (of roughly several to  several tens of parsec in size) embedded in larger structures (a few hundred parsec in size). This feature is inherited from the parental molecular clouds. This is often accompanied by a hierarchical configuration shaped by supersonic turbulence \citep{elmegreen2000,orkisz2017,torniamenti2021}. Many studies have identified  hierarchies in stellar structures in solar neighborhood \citep[e.g.,][]{megeath2016,kounkel2018,ballone2020,kerr2021}. 

Two strategies are commonly used to identify different levels of hierarchical structures. The first strategy is the ``bottom-up'' approach, which initially identifies the lowest-level structures, and merges these to form higher-level larger structures \citep{cantat2019b}. The second strategy is the ``top-down'' method, which starts by initially identifying the highest-level structures and then moves down by fragmentation to locate lower-level substructures \citep{cantat2019a,beccari2018,kerr2021}. 

The Vela OB2 complex is a young stellar grouping located in the direction of the Vela and Puppis constellations, at a distance of 350--400\,pc from the Sun \citep{pozzo2000}. \citet{dezeeuw1999} first made use of proper motions and parallaxes from Hipparcos to identify members of Vela OB2. Subsequently, \citet{pozzo2000} used X-ray observations, and identified a group of pre-main-sequence (PMS) stars in this region, which is often referred to as Gamma Velorum (hereafter Gamma Vel), or Pozzo\,1 \citep{dias2002}. The formation of the Vela OB2 structure is thought to have been triggered by a supernova explosion of a 15\,M$_\odot$ star \citep{cantat2019a} that was located at the center of the IRAS Vela shell \citep{sahu1992}. A similar supernova scenario also has also been suggested for the origin of RCW\,34 complex \citep{bik2010} and the Orion complex \citep{kounkel2020}.
 
Recent studies have identified the hierarchical  clustering in Vela OB2, mostly using ``top-down'' approaches. \citet[][]{beccari2018} adopted DBSCAN method to identify 6 clusters in Vela OB2 covering a distance range of 82\,pc, including Gamma Vel cluster and NGC\,2547. \citet[][]{cantat2019a} used UPMASK algorithm  \citep{cantat2018} and increased the number of identified clusters in Vela OB2 to 11, which extends $\sim$133\,pc in space. Groups in Vela OB2 are likely formed along the densest filaments of the same giant molecular clouds.

Survival of hierarchical clusters in such complexes mainly depends on the star formation efficiency (SFE) and the rate of gas expulsion. The SFE of clusters ranges from a few percent to 30\% or above, depending on the cluster surface density of the region in which the clusters are born \citep{evans2009, megeath2016}. The remaining gas in the cluster is removed through stellar feedback. Violent feedback, such as the feedback caused by supernova explosions, removes the gas much more quickly than  stellar winds or radiation. The faster the gas is expelled, the smaller chance of cluster survival \citep{dinnbier2020a,dinnbier2020b}. Among clusters with an identical gas removal timescale, those with a SFE smaller than 33\% will disrupt faster  \citep{baumgardt2007}.

Besides disruption, another essential physical process in hierarchical clusters in the complex is the mutual gravitational interaction between its substructures. \citet{dela_fuente2009a} suggested that such interactions can be important when the physical separation between neighbouring clusters is less than 30\,pc, which is roughly three times the typical value of the tidal radius for open clusters in the Galactic disk \citep[e.g.,][]{binney2008}. When the relative velocity between two neighbouring clusters is smaller than their internal dispersions \citep{gavagnin2016}, clusters are strongly affected by the  dynamical interaction with their neighbours.
This dynamical interaction may foster the formation of gravitationally-bound cluster pairs \citep[e.g.,][]{priyatikanto2016,arnold2017}. 

Many cluster pairs have been observed in both the Milky Way \citep{dela_fuente2009a,dela_fuente2009b,currie2010}
and in the Large Magellanic Cloud \citep{dieball2002,dieball1998,palma2016}. About half of the Galactic cluster pairs in the work of \citet{dela_fuente2009a,dela_fuente2009b} are younger than 25~Myr, and almost all these pairs are coeval. Since three-dimensional kinematic data was unavailable in earlier studies, determining the physical status of these cluster pairs requires further investigation.

Located a projected distance of 15 degrees from Vela OB2 is the coeval cluster pair of 40\,Myr old, consisting of the clusters Collinder\,135 and UBC\,7 that were first discovered by \cite{castro2018} and were  later studied in further detail by \cite{kovaleva2020}. The distance to this cluster pair is $\sim$300\,pc  \citep{kovaleva2020}. In this work we will investigate the 3D spatial configuration and the kinematic relation between these two constituent clusters.

The Gaia Early Data Release 3 \citep[EDR\,3;][]{gaia2021} provides a 30\% higher precision in parallaxes, and doubled the accuracy in proper motions (PMs), when compared to Data Release 2 \citep[DR\,2;][]{gaia2018}. The 3D spatial positions and PMs allow for an accurate determination of cluster membership, and enables measuring the physical separation between neighboring clusters.
The hierarchical structure of Vela OB2 and the neighboring cluster pair provide an excellent laboratory to study the formation, dynamical evolution, and fate of hierarchical clusters that have formed from the same molecular cloud. Here, we present a detailed study of the Vela OB2 complex, and the cluster pair Collinder\,135 and UBC\,7, using data from Gaia EDR3 and Gaia-ESO survey \citep{gilmore2012}. We aim to quantify the dynamical states of individual clusters and identify interactions between the clusters.  

This paper is organized as follows. In Section~\ref{sec:gaia} we discuss the quality and limitations of the  Gaia EDR\,3 data, and describe our input data-set for member star identification. We then present the algorithm, \texttt{StarGO}, which is used to disentangle the hierarchical clustering in Section~\ref{sec:stargo}. The stellar membership of the hierarchical structures of Vela OB2 and the cluster pair is presented in Section~\ref{sec:membership}, in which the age spread among different clusters in Vela OB2 is discussed. We discuss the kinematic characteristics of each structure in Section~\ref{sec:kinematics}. The 3D spatial configuration and mass distribution of Vela OB2 and the cluster pair are presented in Section~\ref{sec:3D}. In this section, we use the regression discontinuity method  (Section~\ref{sec:mass_stra}) the and minimum spanning tree method (Section~\ref{sec:seg}) to quantify the mass distribution. The dynamical state of the clusters is investigated in Section~\ref{sec:dyn}. We present the star formation history in the target regions in Section~\ref{sec:formation}. In Section~\ref{sec:nbody} we conduct $N$-body simulation to predict the future evolution of Vela OB2 and the cluster pair. Finally, we provide a brief summary of our findings in Section~\ref{sec:summary}.

\section{Identification of hierarchical clustering} \label{sec:member}

\subsection{Gaia EDR3 Data Reduction}\label{sec:gaia}

The analysis described below is carried out for the Vela OB2 complex, and for the cluster pair UBC\,7 and Collinder\,135.
The spatial and kinematic structure of the target regions is investigated using Gaia EDR\,3 data, within 100\,pc of the position of Vela OB2 ($X=-53.9$\,pc, $Y=-334.6$\,pc, $Z=-33.4$\,pc, in heliocentric Cartesian coordinates with the Sun located at the origin), and of the mean position 
between Collinder\,135 and UBC\,7 ($X=-103.7$, $Y=-266.4$\,pc, $Z=-61.3$\,pc), taken from \citet{liu2019}. 
We remove artifacts and poor quality sources from the sample by filtering out stars with relative uncertainties in the parallax or photometry above 10 percent,
and with additional astrometric cuts described in \citet[in their Appendix~C]{lindegren2018}.
After applying this procedure, 327,950 stars remain for subsequent reduction. 

\begin{figure*}[tb!]
\centering
\includegraphics[angle=0, width=1.\textwidth]{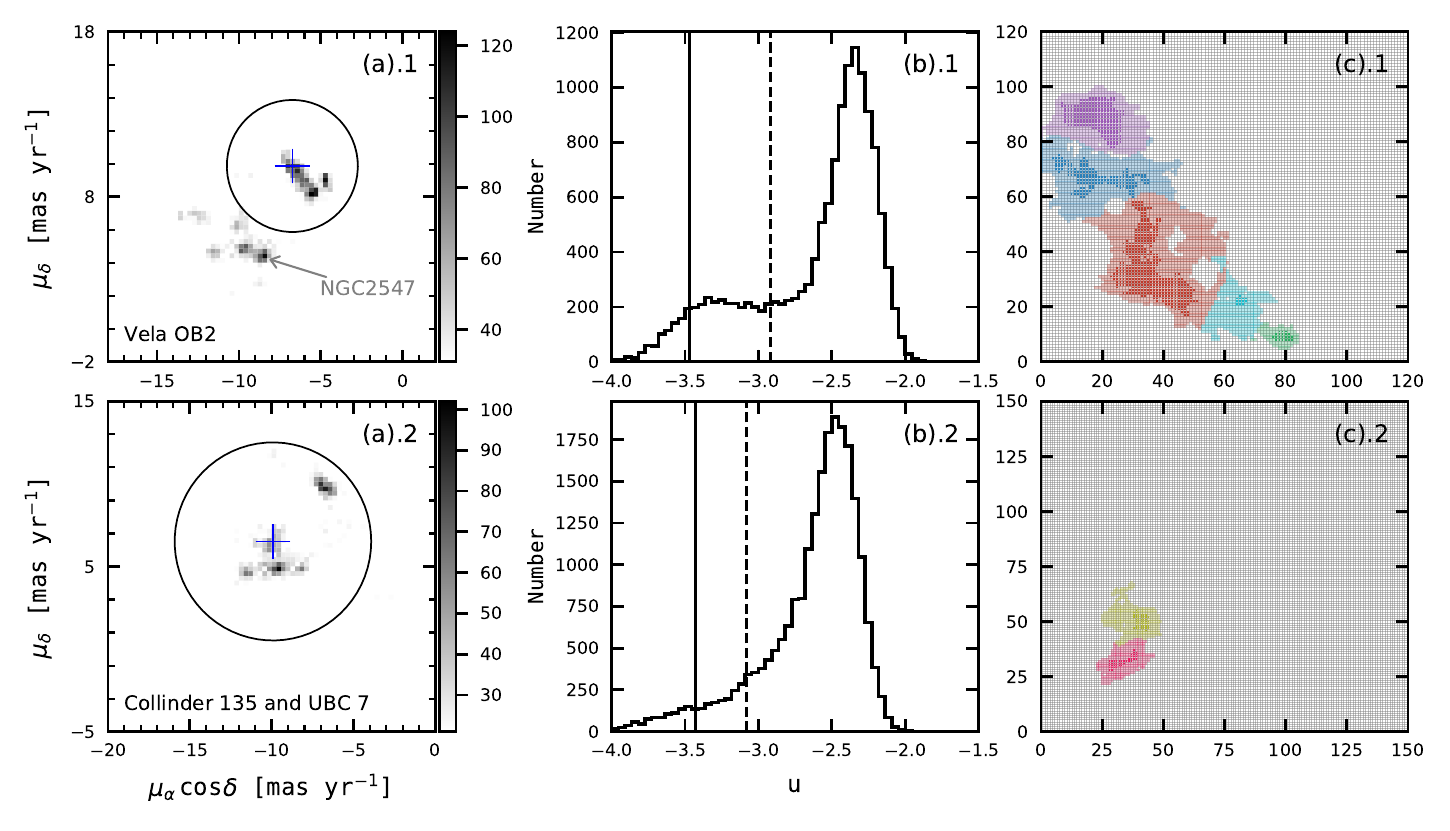}
\caption{ 	 
    (a)~ 2D density map of the proper motion vectors for the regions in Vela OB2 (upper panels) and the cluster pair Collinder\,135 and UBC\,7 (lower panels). The blue crosses indicate the mean over-density of the Gamma Vel cluster and the mean between Collinder\,135 and UBC\,7 taken from \citep{liu2019}. The location of NGC\,2547 is also indicated, which is excluded from the circular proper motion cut. Each bin is smoothed by averaging neighboring eight bins. Only bins with a 
        number count $>3\sigma$ are shown, where $\sigma$ is the standard deviation of all bins.
        The shade of grey indicates the number count in each bin. 
	(b)~ Histogram of the distribution of $u$-matrix elements. 
	     The vertical dashed and solid lines denote the threshold values of $u$ that respectively give a 5\% and a 1\% contamination rate for the identified groups. These correspond to the transparent colored patches and solid colored patches, respectively, 
	    in the 2D neural network (panel (c)). 
	(c)~ 2D neural network resulting from SOM. The neurons with a $u$-threshold corresponding to 5\% contamination rates are shown with transparent colors, which as a whole is considered as the top-level structure. The solid color patches having 1\% contamination rate are the cores of second-level hierarchical substructures. Vela OB2 (the transparent patch) is divided into five  second-level groups, indicated with red, blue, purple, cyan and green. The cluster pair region (transparent patch) is fragmented into Collinder\,135 (yellow) and UBC\,7 (pink). }
\label{fig:som}
\end{figure*}

To allow the algorithm to identify fine structures in the target region more efficiently, we construct a 2D density map of PMs to select stars around the over-densities of Vela OB2 and the cluster pair. 
Figure~\ref{fig:som} panels (a).1 and (a).2 show the 2D density maps of the PMs for the target regions. These maps only show bins with over-densities $>3\sigma$ of the number of stars in all bins. There is a clear over-density near the mean PM \citep{liu2019} of the Gamma Vel cluster. The over-density in the region of Collinder\,135 and UBC\,7 (indicated with a cross) matches the mean PMs of these two clusters measured by \citet[][]{liu2019}. We apply a circular cut (the black circles in Figures~\ref{fig:som}~(a).1 and (a).2) to include the target clusters for further analysis. The radius of the circle is chosen to include as many potential members as possible. Although other over-densities of nearby clusters can be observed in Figure~\ref{fig:som}~(a).1 and (a).2, we only focus on the target clusters and ignore their neighbors in this work.
Application of this circular cut reduces the number of candidate member stars in the two target regions to a total of 31,040. Stars in this sample have magnitudes ranging between $G \sim 3.8$~mag and $G \sim 20.3$~mag. All samples are complete for $G\la 19$--$19.5$~mag. 

In this study we use the 5D parameters of stars after PM selection 
(R.A., Decl., $\varpi$, \pmra, and \pmdec) from Gaia EDR\,3. Radial velocities (RVs) from Gaia DR\,2 are available for only a small fraction of the stars, and have an uncertainty of $\sim$2~km\,s$^{-1}$. 
We adopt the higher-accuracy RVs from \citet{jeffries2014} as supplementary data for the stars in Vela OB2, to aid the study of dynamical state of the clusters. The RVs obtained from \citet{jeffries2014} are part of the Gaia-ESO Survey \citep[GES,][]{gilmore2012}, and have an uncertainty of 0.4~km\,s$^{-1}$.

\subsection{Disentangling Hierarchical Structures}\label{sec:stargo}

\texttt{StarGO}\footnote{\url{https://github.com/zyuan-astro/StarGO-OC}} \citep{yuan2018} has proven to be successful in membership determination of open clusters \citep{tang2019,pang2020,pang2021} and more diffuse stellar streams \citep[e.g.][]{yuan2020a, yuan2020b}. The method is based on the Self-Organizing-Map (SOM), which belongs to the domain of unsupervised-learning. SOM can map high-dimensional data onto a two-dimensional neural network, while preserving the topological structures of the input data. 

We apply \texttt{StarGO} to the selected sample of stars after applying the circular cut in PM space (see Figure~\ref{fig:som} (a).1 and (a).2). The input space to the neural network is of the form ($X, Y, Z$, \pmra, \pmdec). Stars in clusters form over-densities in the input space, and are associated to neuron groups which can be identified by \texttt{StarGO}. We adopt networks of 120$\times$120 and 150$\times$150 neurons (depending on the remaining number of stars after PM selection is applied) for Vela OB2 and the cluster pair, respectively (see Figure~\ref{fig:som}~(c).1 and (c).2). Each neuron is assigned an initially randomized 5D weight vector, which has the same dimensions as the input space. During the learning process, the weight vector of each neuron is updated to become closer to the input vector of a given star. One iteration of the training cycle is finished after the neurons are trained by every star from the entire sample. This process is iterated 400 times when the weight vectors converge. 

After the self-supervised learning is completed, the difference in weight vectors between adjacent neurons is defined as the $u$-matrix. The  distribution of element values in the $u$-matrix is shown in Figure~\ref{fig:som}~(b).1 and (b).2. The tail on the left side of the distribution consists of neurons with relatively low $u$ values, which corresponds to cluster members grouped in the input space. The method for identification is to first associate each star from the entire sample to the neuron that has the closest weight vector to its input vector. Subsequently, we select neurons with $u$ below a threshold value indicated by the dashed line in Figure~\ref{fig:som}~(b).1 and (b).2, which forms the transparent colored patches in panels (c).1 and (c).2. This threshold value of $u$ is selected to ensure an estimated contamination rate of $\sim$5\% among the identified members. Specifically, the contamination rate is estimated from the smooth Galactic disk population using the Gaia EDR\,3 mock catalog \citep{rybizki2020}. An identical PM selection as described in Section~\ref{sec:gaia} is applied to the stars in the mock catalog in the same volume of the sky. Each of these mock stars is attached to the trained 2D neural network. We then count the mock stars associated with selected patches. Assuming that the Milky Way population has the same properties as the mock catalog,  the contamination rate of each identified group can be estimated \citep[see details in][]{pang2020,pang2021} . We adopt this $u$-threshold for the $\sim$5\% contamination rate and obtain the neuron groups shown with transparent colors in Figure~\ref{fig:som}~(c).1 and (c).2.

We obtain one large transparent patch in each of the two runs in the target regions, which represents the top-level structure in Vela OB2 (panel (c).1) and the cluster pair (panel (c).2). To reveal the inner hierarchical substructures, we further decrease  the threshold value of $u$ until the contamination rate reaches 1\%. The corresponding $u$-values are denoted by the solid lines  in Figure~\ref{fig:som} (b).1 and (b).2, and the resulting patches are shown with solid colors in Figure~\ref{fig:som} (c).1 and (c).2. These patches represent the core regions of each cluster.
We then associate neurons in the transparent outskirts to these cores by calculating the minimum difference in  weight vectors between a given neuron and each core-patch. For instance, neurons that have the smallest differences to the red core-patch compared to the other four core-patches are colored red. By this mean, all the neurons in the transparent outskirt are assigned to a core-patch, denoted by different transparent colors in panel (c).1 and (c).2.  In total,  Vela OB2 contains five second-level clusters (five transparent color patches in Figure~\ref{fig:PM} panel (c).1), and the cluster pair is fragmented into 2 second-level structures that correspond to Collinder\,135 and UBC\,7 (the two transparent color patches in Figure~\ref{fig:PM} panel (c).2). This is the first time that the cluster pair Collinder\,135 and UBC\,7 are simultaneously identified with minimal manual intervention. Previous studies \citep{castro2018} did not detect Collinder\,135 using the DBSCAN method since the structure is spatially very extended. 
 
Our identification of hierarchical clustering is purely based on 5D phase-space information without pre-assumption about age, cluster size and density threshold. Therefore, the extended distribution of both young and old generations can be identified independently, which enables us to investigate the progression of star formation inside the stellar complex. Top-level clustering is first extracted at a large scale, which includes significant substructures. We further fragment top-level structures into sub-clusters. This top-down method ensure that the identified hierarchical structures are truly coherent both in space and kinematics. 

\section{Stellar membership in hierarchical structures}\label{sec:membership}
\subsection{Sample Cleaning}\label{sec:clean}

\begin{figure*}[tb!]
\centering
\includegraphics[angle=0, width=1.\textwidth]{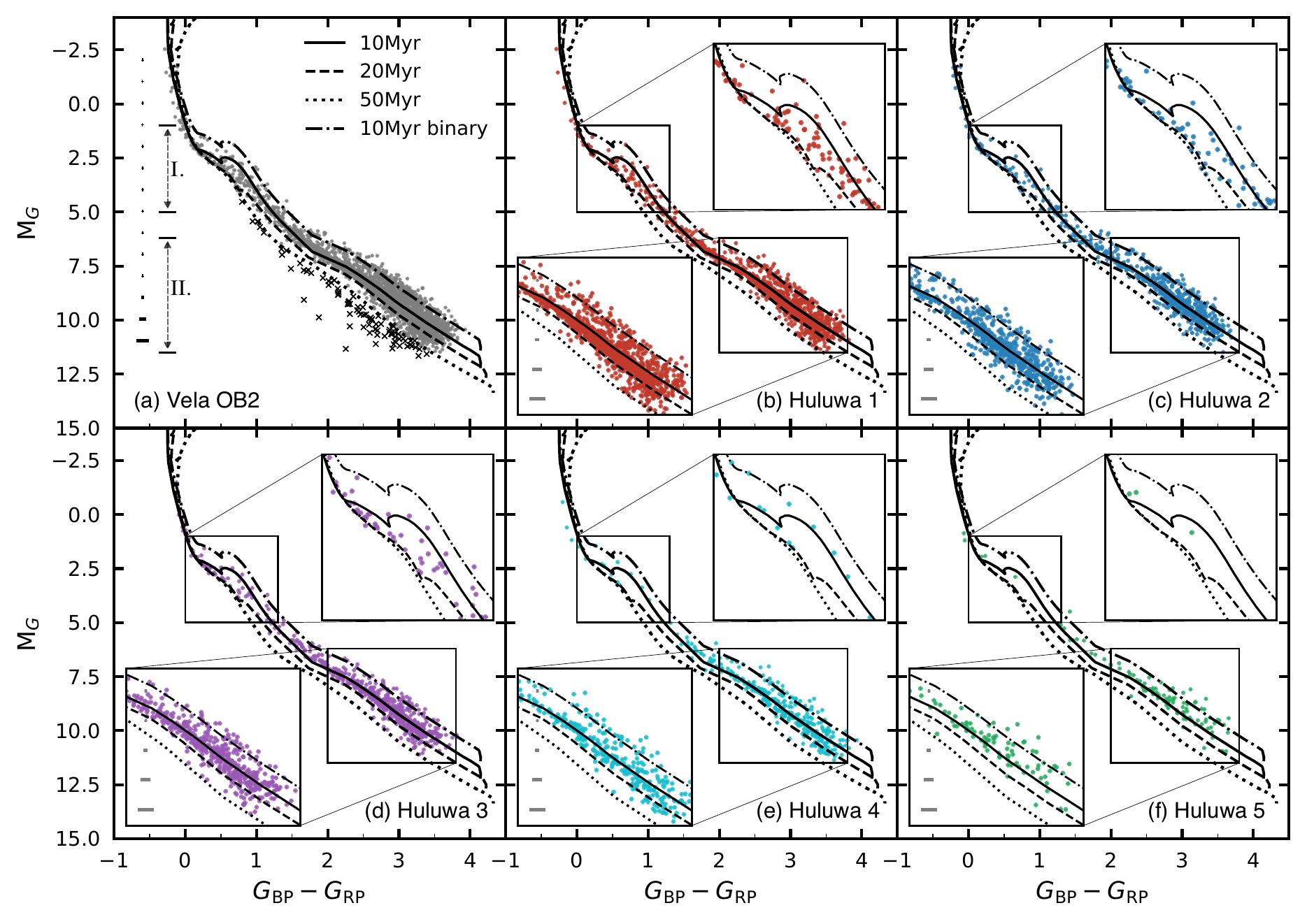}
    \caption{Color-magnitude diagrams for identified structures in Vela OB2. Panel (a): grey dots represent all member candidates of the top-level structure in Vela OB2 (Huluwa\,1--5). Black crosses are field stars. The solid, dashed and dotted curves in (a) are PARSEC isochrones of 10\,Myr, 20\,Myr and 50\,Myr, respectively, with solar metallicity and $A_V=0.35$. The dashed-dotted curve represents the binary sequence of the 10\,Myr isochrone assuming equal mass ratios. The arrows indicate the upper PMS (I) and lower PMS (II) regions. The format of the curves in panels (b)--(f) are identical to those in panel (a). The two insets show zoomed-in versions of the upper PMS (I) and lower PMS (II) regions.}
\label{fig:cmd1}
\end{figure*}

\begin{figure*}[tb!]
\centering

\includegraphics[angle=0, width=1.\textwidth]{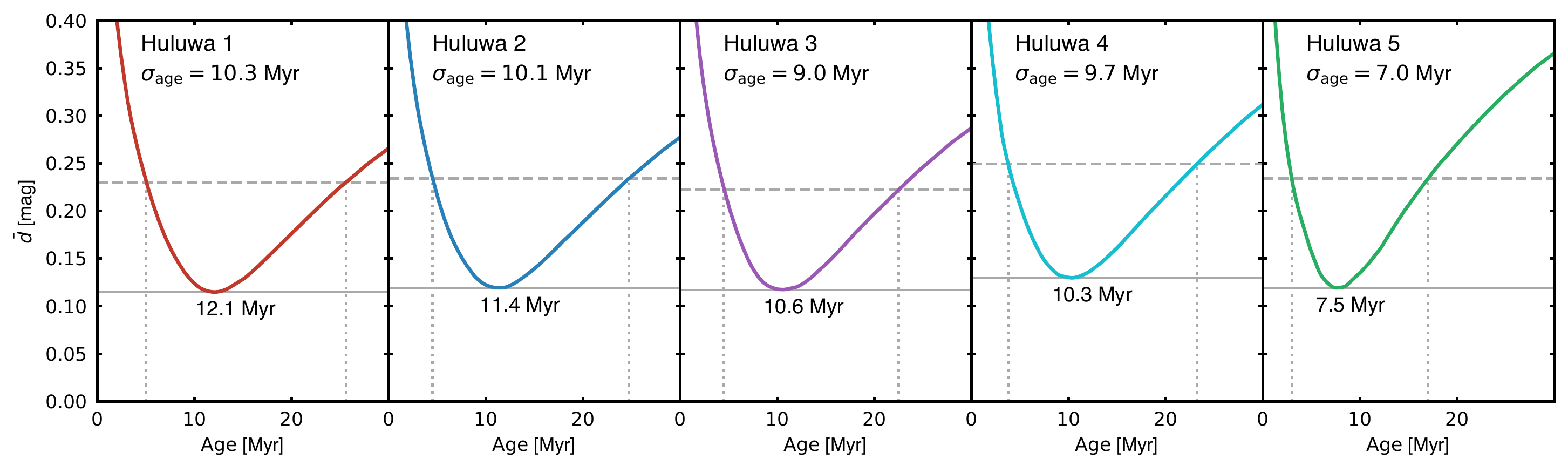}
    \caption{
    Dependence of the average distance $\bar{d}$ on the cluster age. Colored curves show the average distance $\bar{d}$ between members of each group and an isochrone at a given age as measured in the color-magnitude diagram. The solid grey horizontal line indicates the minimum average distance at the best fitted age. The horizontal grey dashed line corresponds to the 1\,$\sigma$ deviation above the minimum average distance. The uncertainty of the fitted age is taken as half of the age difference between two vertical dotted grey lines. }
\label{fig:age_scat}
\end{figure*}

\begin{figure}[tb!]
\centering
\includegraphics[angle=0, width=1.0\columnwidth]{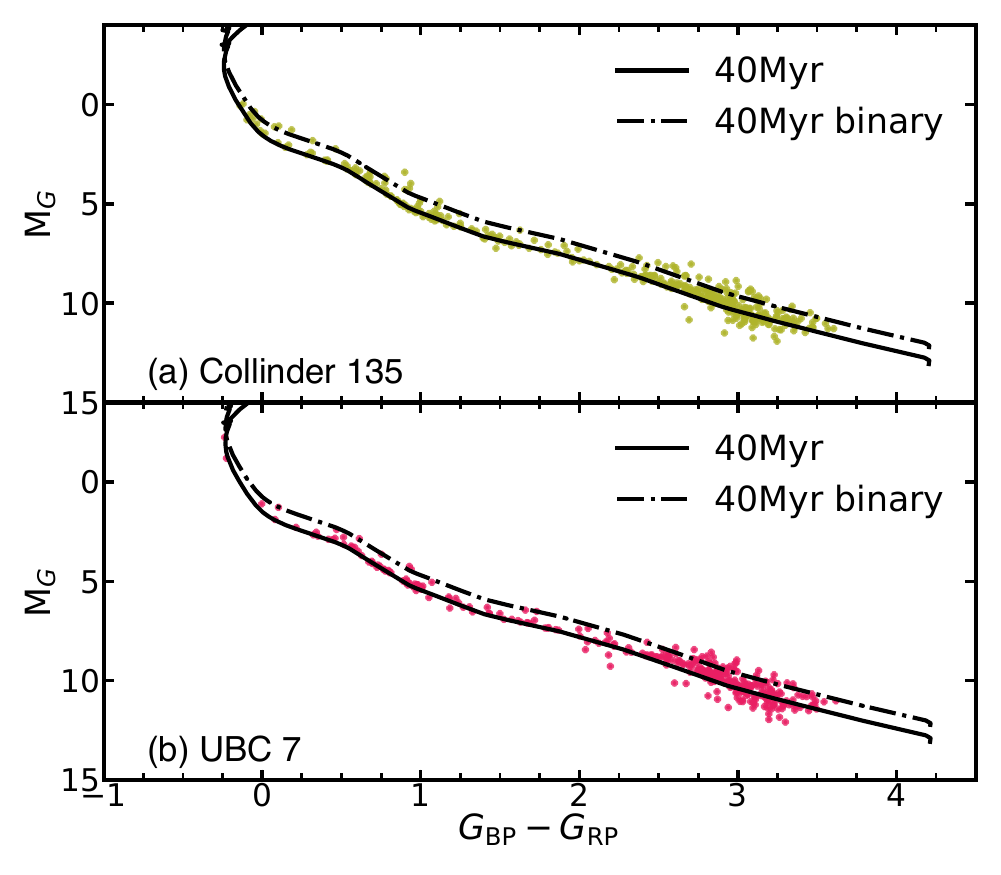}
    \caption{Color-magnitude diagrams for Collinder\,135 and UBC\,7. A PARSEC isochrone of 40\,Myr (black solid curve) is displayed in panels (g) and (h) with solar abundance and $A_V=0.13$. The dashed-dotted curve represents the binary sequence of the 40\,Myr isochrone, assuming equal mass ratios.}
\label{fig:cmd2}
\end{figure}

We construct color-magnitude diagrams (CMDs) for the member candidates of the five second-level clusters in Vela OB2 (Figure~\ref{fig:cmd1}), and for Collinder\,135 and UBC\,7 (Figure~\ref{fig:cmd2}). Stars in the top-level structure of Vela OB2, together (grey dots in Figure~\ref{fig:cmd1} panel (a)), track a locus of a main sequence (MS). Previous estimated age of Vela OB2 ranges from 10\,Myr via isochrone fitting \citep{prisinzano2016,cantat2019a} to 18--21\,Myr via Lithium abundance \citep{jeffries2017} . We plot PARSEC isochrones of 10\,Myr and 20\,Myr (black solid and dashed curves) with solar metallicity and $A_V=0.35$ \citep{cantat2019a}. 
A large number of stars are still in the PMS stage, while less are located in the MS. A significant number of upper PMS stars are older than 10\,Myr, i.e., bluer than the 10\,Myr-old isochrone with M$_G$ in the 2.0--5.0\,mag range (region I) in Figure~\ref{fig:cmd1} (a)), which were excluded from \citet{cantat2019a}, as they excluded the older population from NGC\,2547. NGC\,2547 has been excluded from our analysis using the circular cut of PMs (see Figure~\ref{fig:som} (a).1). The older upper PMS stars in our sample therefore belong to the Vela OB2 complex. The blue edge of the upper PMS (region I) follows the 20\,Myr isochrone. We overplot a binary sequence of the 10\,Myr isochrone with mass ratio equal to unity (dashed-dotted curve). With binary stars and photometric uncertainties taken into account, the broadening of the PMS at region I and II may still imply an age spread in the order of a million years in Vela OB2. Such age spread is commonly seen in other young star clusters \citep[e.g., NGC\,3603,][]{beccari2010,pang2013} as a consequence of the propagation of star formation.

A handful of candidate stars are located below the MS locus (M$_G>4.3$\,mag), and correspond to an age older than 50\,Myr (black dotted curve in Figure~\ref{fig:cmd1}~(a)). These stars are located in a more extended region than the majority of the stars in Vela OB2. These stars are likely field stars. As the cluster is located very close to the Galactic plane, field star contamination is significant, and additional photometric cleaning is required \citep[see Section 2.3 in ][]{pang2020}.
Therefore, we consider stars bluer than the 50\,Myr-old isochrone and fainter than M$_G>4.3$\,mag as field stars (black crossed symbols in panel (a) in Figure~\ref{fig:cmd1}), and exclude them from the analysis below. 

We label the five second-level clusters in Vela OB2 as Huluwa\footnote{Huluwa means calabash brothers, which is one of the most popular animations in China in the 1980s. The legend begins when the calabashes sequentially fall off from the same stem, and transform into seven boys (\url{https://en.wikipedia.org/wiki/Calabash_Brothers}). Similarly, the five clusters are sequentially born from the same molecular cloud.} \,1--5, following a Chinese animation series, Huluwa (Calabash brothers). From Huluwa\,1 to Huluwa\,5, the number of stellar members decreases. Huluwa\,1 is the most populous group, which is the Gamma Vel (Pozzo\,1) cluster.

\subsection{Age spread in the hierarchical structures}\label{sec:age}

We construct the CMD for each group in Vela OB2 after cleaning additional contamination in panels (b)--(f) in Figure~\ref{fig:cmd1}. In order to identify a possible age difference between the five clusters, we enhance the upper PMS region (I) and lower PMS region (II) in the CMD (indicated with arrows in panel (a)). The former is the region where the PMS merges with the MS, and the latter is a region where a significant spread exists. The PMS isochrones are not well established in the lower PMS region \citep{kos2019}. Previous studies have shown that coeval stars with different accretion rates will induce a spread in luminosity, especially in the lower-mass end of the PMS \citep{tout1999, baraffe2009, kunitomo2017}. As can be seen in Figure~\ref{fig:cmd1} (b--f), stars in the lower PMS region spread out to the 10\,Myr binary sequence in all groups. Based on the above considerations, the age estimation from this lower PMS region will suffer from large uncertainties.

Stars in the upper PMS regions in Huluwa\,1--3 populate the region between the 10 and 20\,Myr isochrones (panels b--d), suggesting that these structures have ages between 10\,Myr and 20\,Myr.
On the contrary, stars in the upper PMS region in Huluwa\,4 and Huluwa\,5 are mostly younger than the 10\,Myr isochrone. However, these two clusters suffer from low number statistics in this region.

To further quantify the age spread among Huluwa\,1--5, we adopt and modify the method used in \citet{liu2019} and \citet{kos2019}. We define the average distance of each member in the cluster to the isochrone in the CMD as
\begin{equation}\label{equ:2}
    \bar{d}=\frac{1}{n}\sum_{k=1}^{n}\sqrt{[\Delta (G_{\rm BP}-G_{\rm RP})_{i,n}]^2+[\Delta {\rm M_G}_{i,n}]^2}
\end{equation}
where $\Delta (G_{\rm BP}-G_{\rm RP})_{i,n}$ and $\Delta {\rm M_G}_{i,n}$ are respectively the color difference and magnitude difference between the member star to its nearest point of the isochrone with the $k$-D tree method \citep{millman2011}. We study a set of isochrones with solar metallicity, $A_V=0.35$ \citep{cantat2019a}, and age ranging from 0\,Myr to 30\,Myr with a step size of 0.1\,Myr. The isochrone with the smallest value of $\bar{d}$ (solid grey line in Figure~\ref{fig:age_scat}) is considered as the best fitted age of the cluster. The value of $\bar{d}$ shows a local minimum and increases towards older and younger ages (Figure~\ref{fig:age_scat}). 
The uncertainty of the fitted age $\sigma_{\rm age}$ of each group is taken as half of the age difference (between the two vertical grey lines) at which the average distance reaches 1\,$\sigma$ (grey dashed line) above the minimum $\bar{d}$.

The age uncertainty is comparable to the best fitted age, which further confirms an existence of an age spread found in Figure~\ref{fig:cmd1}. Although an absolute age determination for young clusters using isochrone fitting may have an uncertainty of 10\% \citep{kos2019}, a relative age spread of a few, up to $\sim 10$\,Myr in Huluwa\,1--5 still can be identified in Figure~\ref{fig:age_scat}. Note that the effects of stellar binarity, the accretion of PMS, and the photometric uncertainty can also position stars above or below the isochrone and contribute to the average distance. Based on our current data, we are unable to disentangle these three contributions. Therefore, the estimated age spread from the uncertainty $\sigma_{\rm age}$ should be treated as an upper limit for the age spread in each group. 
 
From Huluwa\,1 to Huluwa\,5, the age becomes younger and the age spread decreases. A progression of star formation could have taken place in Huluwa\,1--5.  
The best fitted age of Huluwa\,1 is 12.1~Myr with a spread of 10.3\,Myr. 
Huluwa\,5 is the youngest cluster with an age of 7.5\,Myr and the smallest age spread of 7\,Myr.  
This age estimation is consistent with the previous findings of \citet{jeffries2017}, who suggested an age 18--21\,Myr for Gamma Vel (Huluwa\,1) based on lithium-abundance analysis, and is also consistent with the recent work by \citet{franciosini2020}, who used a lithium-depletion model of 18\,Myr with 20\% spot coverage and achieved a reasonable fit for Gamma Vel (Huluwa\,1).

 Considering the age spread in each group, we adopt an age of 10\,Myr for Huluwa\,1--5 in Vela OB2, and obtain the mass of each member star from the nearest point on the isochrone via the $k$-D tree method \citep{millman2011}. The most massive group, Huluwa\,1 contains 723.5\,M$_\odot$. Huluwa\,2 is the second most massive cluster with 467.4\,M$_\odot$. The total mass for the 5 clusters in Vela OB2 is 1805\,M$_\odot$, with an uncertainty of 104\,M$_\odot$ when a different age of 20\,Myr is taken into consideration. 
 
The CMD locations of the members of cluster pair, Collinder\,135 and UBC\,7 (Figure~\ref{fig:cmd2}) are consistent with an age of 40\,Myr \citep{kovaleva2020}. 
 A binary sequence is clearly visible in the CMDs of both Collinder\,135 and UBC\,7, which is within the equal-mass binary sequence  (dashed-dotted curves). However, no obvious contamination sequence is observed. We do not apply additional cleaning for these two clusters. 
Adopting an age of 40\,Myr for each of the clusters in the pair, the total mass of the cluster pair reaches 445.5\,M$_\odot$ (Collinder\,135: 253.2\,M$_\odot$ and UBC\,7: 192.3\,M$_\odot$).

\subsection{Membership Cross-match}\label{sec:match}
 
\begin{deluxetable*}{lc RRR RR r RR rrr c}
\tablecaption{General parameters of target clusters \label{tab:general}
             }
\tabletypesize{\scriptsize}
\tablehead{
	 \colhead{Cluster}          & \colhead{Age}         & 
	 \colhead{$X_c$}            & \colhead{$Y_c$}       & \colhead{$Z_c$}           & 
     \colhead{$r_{\rm h}$}      & \colhead{$r_{\rm t}$} & \colhead{$M_{\rm cl}$}    &
	 \colhead{\pmra}            & \colhead{\pmdec}      &
	 \colhead{$\sigma_{\rm PMRA}$}   & \colhead{$\sigma_{\rm DEC}$}   & \colhead{$\sigma_{\rm RV}$} & 
	 \colhead{memb.}        \\
	 \colhead{}                 & \colhead{(Myr)}       &
	 \multicolumn{5}{c}{(pc)}   & \colhead{(M$_\sun$)}  &  
	 \multicolumn{4}{c}{(mas yr$^{-1}$)} &
	 \colhead{(km\,s$^{-1}$)}   & \colhead{(number)}    \\
	 \cline{3-7} \cline{9-12}
	 \colhead{(1)} & \colhead{(2)} & \colhead{(3)} & \colhead{(4)} & \colhead{(5)} &
	 \colhead{(6)} & \colhead{(7)} & \colhead{(8)} & \colhead{(9)} & \colhead{(10)} &
	 \colhead{(11)} & \colhead{(12)} & \colhead{(13)} & \colhead{(14)}
	 }
\startdata
Huluwa\,1           & 12.1--22.4    & -46.3     & -348.6 & -46.1 & 14.6 & 13.1 & 724    & -6.38     &  9.33 & 0.62$^{+0.07}_{-0.05}$  & 1.01$\pm$0.10            & 1.34$^{+1.01}_{-0.84}$  & 1294 \\
\ \ (Huluwa\,1A)    & 12.1--22.4    & -44.2     & -342.6 & -45.7 & 10.8 & 10.6 & 381.2  & -6.55     &  9.72 & 0.46$\pm$0.03           & 0.39$\pm$0.03            & 0.35$^{+0.16}_{-0.20}$  & 675  \\
\ \ (Huluwa\,1B)    & 12.1--22.4    & -48.2     & -354.9 & -47.0 & 13.6 & 10.2 & 342.8  & -6.20     &  8.90 & 0.49$\pm$0.07           & 0.38$\pm$0.05            & 0.92$^{+0.41}_{-0.49}$  & 619  \\
Huluwa\,2           & 11.4--21.5    & -45.4     & -391.2 & -63.2 & 15.3	& 11.3 & 467    & -5.55     &  8.22 & 0.81$^{+0.11}_{-0.09}$  & 0.66$\pm$0.08            & 2.35$^{+1.08}_{-0.92}$  & 743  \\
Huluwa\,3           & 10.6--19.6    & -64.2     & -383.1 & -70.1 &  7.9	& 10.5 & 373    & -4.74     &  8.94 & 0.77$^{+0.17}_{-0.15}$  & 0.75$^{+0.13}_{-0.11}$   & 1.80$^{+1.20}_{-0.85}$  & 588  \\
Huluwa\,4           & 10.3--20.0       & -64.6     & -335.4 & -12.3 & 14.8	&  8.3 & 181    & -7.15     & 10.02 & 0.68$^{+0.24}_{-0.16}$  & 0.50$^{+0.18}_{-0.11}$   & \nodata              & 347   \\
Huluwa\,5           & 7.5--14.5        & -95.4     & -341.7 &  12.8 &  4.7	&  5.7 &  61    & -7.01     & 10.81 & 0.49$^{+0.34}_{-0.15}$  & 0.47$^{+0.30}_{-0.15}$   & \nodata              & 102   \\
Collinder\,135      & 40        & -105.8    & -277.7 & -58.0 &  9.7	&  9.2 & 253    & -10.05    &  6.17 & 0.73$^{+0.10}_{-0.09}$  & 0.69$^{+0.09}_{-0.07}$   & 1.13$\pm$0.53          & 377   \\
UBC\,7	            & 40        & -98.0     & -252.6 & -63.7 &  7.6	&  8.4 & 192    & -9.77     &  7.00 & 0.46$^{+0.16}_{-0.12}$  & 0.51$^{+0.07}_{-0.05}$   & 1.36$^{+0.72}_{-0.76}$ & 336   \\
\enddata
\tablecomments{Note that Huluwa\,1A \& 1B are two components of Huluwa\,1. 
$X_c$, $Y_c$, and $Z_c$ are the median position of members in each cluster, in heliocentric Cartesian coordinates after distance correction (Section~\ref{sec:space}). $r_{\rm h}$ and $r_t$ are half-mass and tidal radii of each cluster, respectively. The tidal radius is computed using equation 12 in \citet{pinfield1998}. The quantity $M_{\rm cl}$ is the total mass of each star cluster. \pmra ~and \pmdec ~are the mean PMs for each group.  $\sigma_{\rm pmra}$ and $\sigma_{\rm pmdec}$ are the dispersions of the R.A. and Decl. components of the PMs. $\sigma_{\rm RV}$ is the RV dispersion after correction of binary stars and field star contamination. The latter is not available for Huluwa\,4 and 5 due to the small number of RV measurements. Errors in the velocity dispersion are obtained using the MCMC method.}
\end{deluxetable*}

In total, 3074 members of Vela OB2 are selected after CMD cleaning, 1294 members for Huluwa\,1, 743 members for Huluwa 2, 588 members for Huluwa\,3, 347 members for Huluwa\,4, and 102 members for Huluwa\,5. Collinder\,135 contains 377 members and 336 members are identified for UBC\,7. Parameters of each of these structures are listed in Table~\ref{tab:general}. We provide a detailed member list of all clusters in Table~\ref{tab:memberlist}, with parameters obtained in this study. 

Our member list reaches a good agreement with those of earlier studies, but provides a more complete sample for Vela OB2. For the largest group, Huluwa\,1, we extend the number of members from 437 in \citet{liu2019} (ID:2394, also known as LP\,2394) and 341 in \citet{cantat2020} to 1294, where 354 and 325 members are overlapped with the current study, respectively. There are 139 members in Huluwa\,1 and 27 members in Huluwa\,2 that cross-match with members identified in \citet{jeffries2014}. 

Among the members of the cluster pair, we compare our member list to that of \citet{cantat2020}, in which 232 stars matched in Collinder\,135, and 118 in UBC\,7. 
 
\begin{deluxetable*}{ccc}
\tablecaption{Columns for the table of individual members of the five clusters in Vela OB2, and the cluster pair Collinder\,135 and UBC\,7. \label{tab:memberlist}
             }
\tabletypesize{\scriptsize}
\tablehead{
\colhead{Column}    & \colhead{Unit}    & \colhead{Description}
}
\startdata
Cluster Name                    &                  &  Name of the target cluster   \\
      Gaia ID                   &                  &  Object ID in Gaia EDR\,3\\
ra                              & degree           &  R.A. at J2016.0 from Gaia EDR\,3\\
er\_RA                          & mas              &  Positional uncertainty in R.A. at J2016.0 from Gaia EDR\,3 \\
dec                             & degree           &  Decl. at J2016.0 from Gaia EDR\,3 \\
er\_DEC                         & mas              &  Positional uncertainty in decl. at J2016.0 from Gaia EDR\,3 \\
parallax                        & mas              &  Parallax from Gaia EDR\,3\\
er\_parallax                    & mas              &  Uncertainty in the parallax \\
pmra                            & mas~yr$^{-1}$    &  Proper motion with robust fit in $\alpha \cos\delta$ from {\it Gaia} EDR\,3     \\
er\_pmra                        & mas~yr$^{-1}$    &  Error of the proper motion with robust fit in $\alpha \cos\delta$   \\
pmdec                           & mas~yr$^{-1}$    &  Proper motion with robust fit in $\delta$ from Gaia EDR\,3     \\
er\_pmdec                       & mas~yr$^{-1}$    &  Error of the proper motion with robust fit in $\delta$  \\
Gmag                            & mag              & Magnitude in $G$ band from Gaia EDR\,3   \\
BR                              & mag              & Magnitude in $BR$ band from Gaia EDR\,3   \\
RP                              & mag              & Magnitude in $RP$ band from Gaia EDR\,3   \\
Gaia\_radial\_velocity          & km~s$^{-1}$      &  Radial velocity from Gaia DR\,2 \\
er\_Gaia\_radial\_velocity      & km~s$^{-1}$      &  Error of radial velocity from Gaia EDR\,3\\
Jeffries\_radial\_velocity       & km~s$^{-1}$      &  Radial velocity from Gaia/ESO survey \citep{jeffries2014}\\
er\_Jeffries\_radial\_velocity   & km~s$^{-1}$      &  Error of radial velocity from Gaia/ESO survey \citep{jeffries2014}\\
Mass                            & M$_\odot$        & Stellar mass obtained in this study\\
X\_obs                          & pc               & Heliocentric Cartesian X coordinate computed via direct inverting Gaia EDR\,3 parallax $\varpi$ \\
Y\_obs                          & pc               & Heliocentric Cartesian Y coordinate computed via direct inverting Gaia EDR\,3 parallax $\varpi$ \\
Z\_obs                          & pc               & Heliocentric Cartesian Z coordinate computed via direct inverting Gaia EDR\,3 parallax $\varpi$ \\
X\_cor                          & pc               & Heliocentric Cartesian X coordinate after distance correction in this study \\
Y\_cor                          & pc               & Heliocentric Cartesian Y coordinate after distance correction in this study \\
Z\_cor                          & pc               & Heliocentric Cartesian Z coordinate after distance correction in this study \\
Dist\_cor                       & pc               & The corrected distance of individual member\\
\enddata
\tablecomments{A machine readable full version of this table is available online.}
\end{deluxetable*}

\section{Kinematical characteristics of the hierarchical structures}\label{sec:kinematics}

\subsection{Proper Motion Distribution}\label{sec:PMD}
 
\begin{figure*}[tbh!]
\centering
\includegraphics[angle=0, width=1.\textwidth]{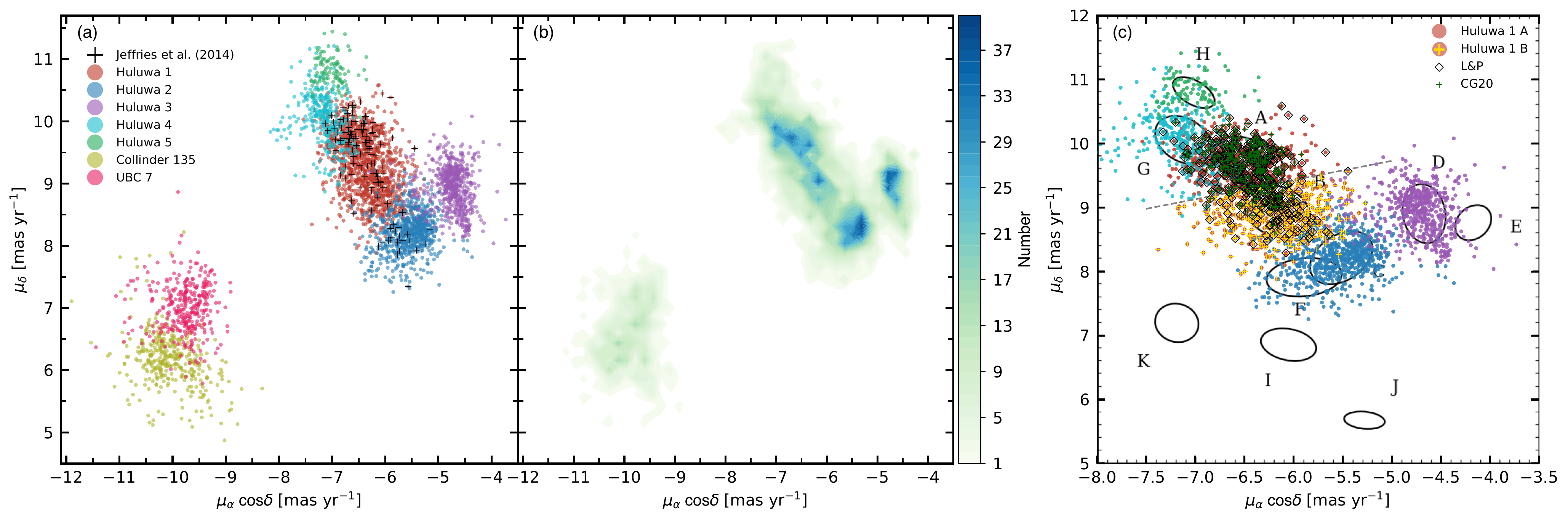}
    \caption{(a): proper motion diagram for the five clusters in Vela OB2 and the cluster pair Collinder\,135 and UBC\,7. Colored dots indicate members 
                    of different groups. Black crosses indicate members that have GES RV measurements from \citet{jeffries2014}. 
                    (b): contour map of the proper motion distribution.  
            (c): modified proper motion diagram of the top-left panel in Figure~5 of \citet{cantat2019a}. 
                    Members of Huluwa\,1--5 are over-plotted. Each ellipse represents the mean and standard deviation of 11 groups in \citet{cantat2019a}. The dotted straight line divides Huluwa\,1 into the two subgroups Huluwa\,1A (red dots) and Huluwa\,1B (yellow crosses). Members of Huluwa\,1 that are cross-matched with the catalogues of  \citet{liu2019} and \citet{cantat2020} are indicated as black diamonds and dark-green 
                   crosses, respectively. 
            }
\label{fig:PM}
\end{figure*}

One characteristic of hierarchical structures in the complex is kinematic coherence between the stellar groups. Such coherence is expected since all substructures were formed from the same molecular cloud, and have inherited the internal cloud kinematics \citep[see, e.g.,][]{elmegreen2000}. The PM distribution and contour maps of Huluwa\,1--5 and the cluster pair are presented in panels (a) and (b) of Figure~\ref{fig:PM}. Each cluster occupies a distinct but adjacent region in the PM vector diagram, suggesting that their kinematical properties are related. Huluwa\,1 exhibits an extended PM distribution. Two over-densities are observed within Huluwa\,1 in the PM contour map (see Figure~\ref{fig:PM} (b)). The PM distributions of Collinder\,135 and UBC\,7 are adjacent but still well-separated, which supports a common formation origin for this coeval cluster pair \citep{castro2018}. 

Previous studies have succeeded in disentangling Vela OB2 into different hierarchical groups. In panel (c) of Figure~\ref{fig:PM}, we overplot the location of the 11 groups described in \citet[][CG19]{cantat2019a} in the PM distribution. Huluwa\,1--5 overlap with 7 groups from CG19. Huluwa\,1 includes groups~A and~B of \citet{cantat2019a}, which is consistent with the PM contour over-density in panel (b).
Groups~I, J, K of CG19 are more distant structures (more than 100\,pc from Vela OB2) that are not included in our analysis. Additionally, Huluwa\,1--3 relate to groups~4, 2, and 1 in \citet[][]{beccari2018}. Note that 27 members in the Gamma Vel B component (Huluwa\,1) from 
 \citet[][black crosses in the blue-dot region in panel (a)]{jeffries2014} do not actually belong to Gamma Vel (Huluwa\,1). Instead they are members of another cluster Huluwa\,2 (blue dots in panel (a)).

We follow the PM locations of groups A and B in \citet{cantat2019a}, and separate Huluwa\,1 into Huluwa\,1A and Huluwa\,1B for the following kinematic analysis, with a straight line in panel (c) of Figure~\ref{fig:PM}. Compared to previous studies, cross-matched members of Huluwa\,1 (Gamma Vel) with \citet[][black diamonds]{liu2019} have a larger coverage in Huluwa\,1B than \citet[][dark green crosses]{cantat2020}. The adjacent distribution and the closeness of PMs in Huluwa\,1A and Huluwa\,1B resembles the configuration of the cluster pair Collinder\,135 and UBC\,7.

\subsection{Radial Velocity Distributions}

\begin{figure}[th!]
\centering
\includegraphics[angle=0, width=1 \columnwidth]{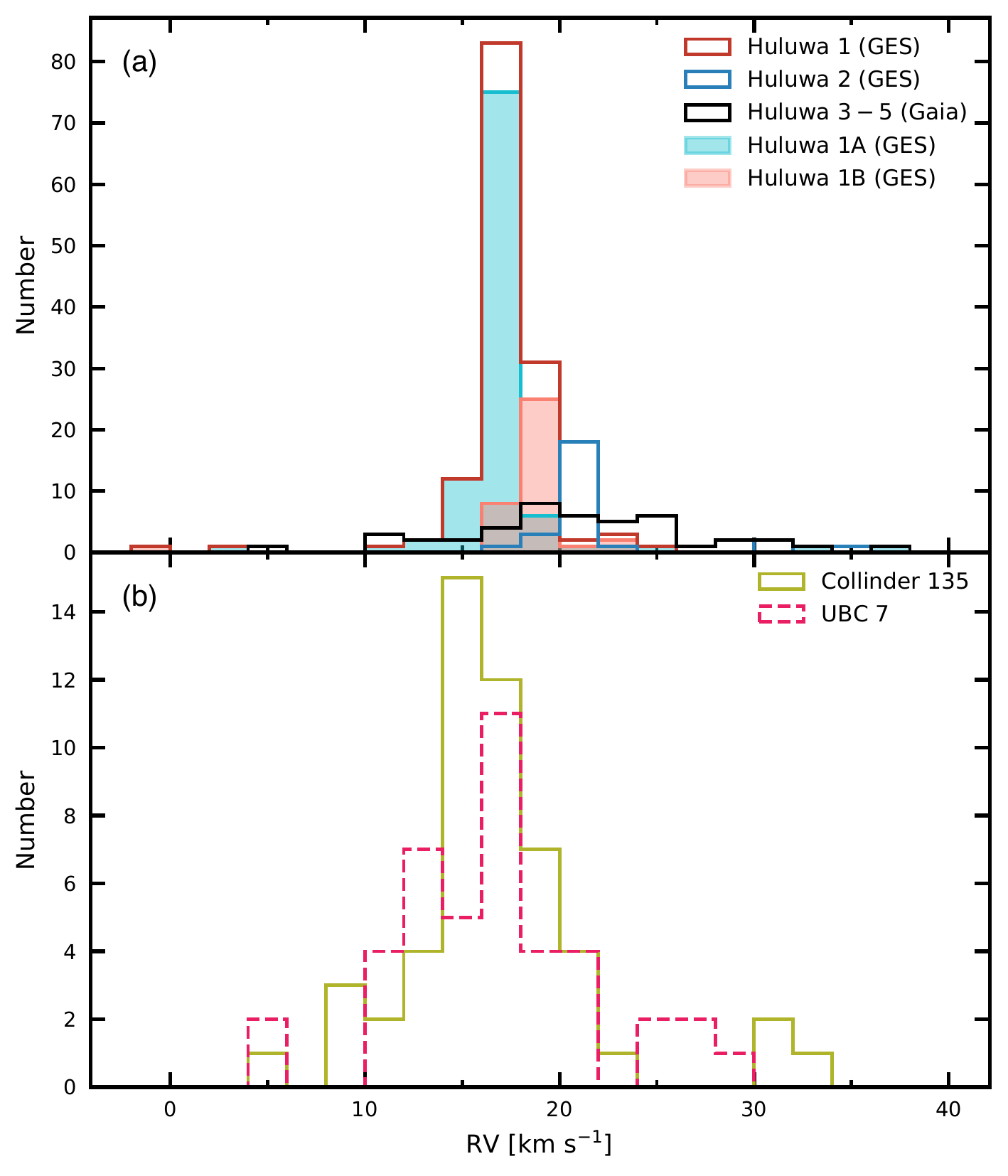}
    \caption{
        Histograms of radial velocities for Huluwa\,1--5 in Vela OB2 (a), and the cluster pair Collinder\,135 and UBC\,7 (b). 
        The bin size of the histogram is 2\,km~s$^{-1}$. 
        Histograms for Huluwa\,1 (red) and Huluwa\,2 (blue) are obtained from members with GES RVs from \citet{jeffries2014} only. We separate Huluwa\,1 into Huluwa\,1A (cyan-shaded histogram) and Huluwa\,1B (salmon-shaded histogram) following the division criteria in Figure~\ref{fig:PM} (c). RVs of Huluwa\,3--5 and the cluster pair were obtained from Gaia DR\,2. }
\label{fig:rv}
\end{figure}

Earlier studies have found that the most massive cluster, Huluwa\,1 (Gamma Vel), features two kinematically distinct coeval stellar populations \citep{jeffries2014,sacco2015,franciosini2018}, which were suggested to correspond to a bound core and an expanding halo \citep{jeffries2014}, or they may have been generated by a past encounter between Gamma Vel and the nearby cluster NGC\,2547 \citep{damiani2017}. In previous studies, the member list of Gamma Vel B (Huluwa\,1B) included members of Huluwa\,2 (see Figure~\ref{fig:PM} (a) black crosses); this may result in an overestimation of the computed velocity dispersion.
 We therefore re-build the RV distribution of Huluwa\,1 using RV measurements from GES (Figure~\ref{fig:rv} (a)), based on members identified in the current paper. 
Huluwa\,1A ($\sim$17.1\,km\,s$^{-1}$; shaded cyan histogram) is distinct from Huluwa\,1B ($\sim$18.2\,km\,s$^{-1}$; shaded salmon histogram) with an RV distribution of mean RV offset by 1\,km\,s$^{-1}$, which supports the earlier claims of two distinct kinematic components in Huluwa\,1.

The RV distribution in Huluwa\,2 (blue histogram, using RV from GES) generally is more dispersed, with a single peak at 21.0\,km\,s$^{-1}$. For Huluwa\,3--5 (black histogram, using RV from Gaia DR2), the RV profile is much more extended.  Note that the number of RV measurements in Huluwa\,3--5 is much smaller, and the RV error in Gaia DR2 is roughly a factor ten higher than that of GES.
For the cluster pair, Collinder\,135 has an almost identical RV distribution as UBC\,7 (Figure~\ref{fig:rv} (b)), with a mean RV of 16.7\,km\,s$^{-1}$ and 17.3\,km\,s$^{-1}$ respectively. The RV similarity between the cluster pair constituents is very similar to that of Huluwa\,1A and Huluwa\,1B. In both Vela OB2 and the cluster pair, we observe stars with RVs that deviate from the bulk of the population; these are likely stellar binary systems.
Since RV information is not included in our member identification, we do not recover this two third-level structures in Huluwa\,1 directly from \texttt{StarGo}. 
Note that groups~A and~B are identified in different parallax ranges in \citet{cantat2019a}.

\section{3D space and mass distribution}\label{sec:3D}
\subsection{Spatial Configuration}\label{sec:space}

We construct the 3D spatial distribution for each cluster by correcting the positions of individual stars for the pseudo-elongation generated by parallax errors, via the Bayesian method described in \citet{carrera2019} and \citet{pang2020,pang2021}. We adopt a Gaussian distribution for the distances to the cluster stars (for extended structures, a Gaussian profile only provides a rough estimation) and an exponential profile \citep{bailer2015} for the distances to field stars in the prior.
We adopt the median value of cluster-centric distance of individual stars as the scale radius of the Gaussian distribution. 
The membership probability of each star is 95\%, considering a 5\% field star contamination rate resulting from member identification (see Section~\ref{sec:stargo}).
All members are located at a distance of $278-399$\,pc. Therefore, the uncertainty in the corrected distance for stars is 2--3\,pc, according to the simulations in \citet[see their Figure 4]{pang2021}. 

\begin{figure*}[tb!]
\centering
\includegraphics[angle=0, width=0.95\textwidth]{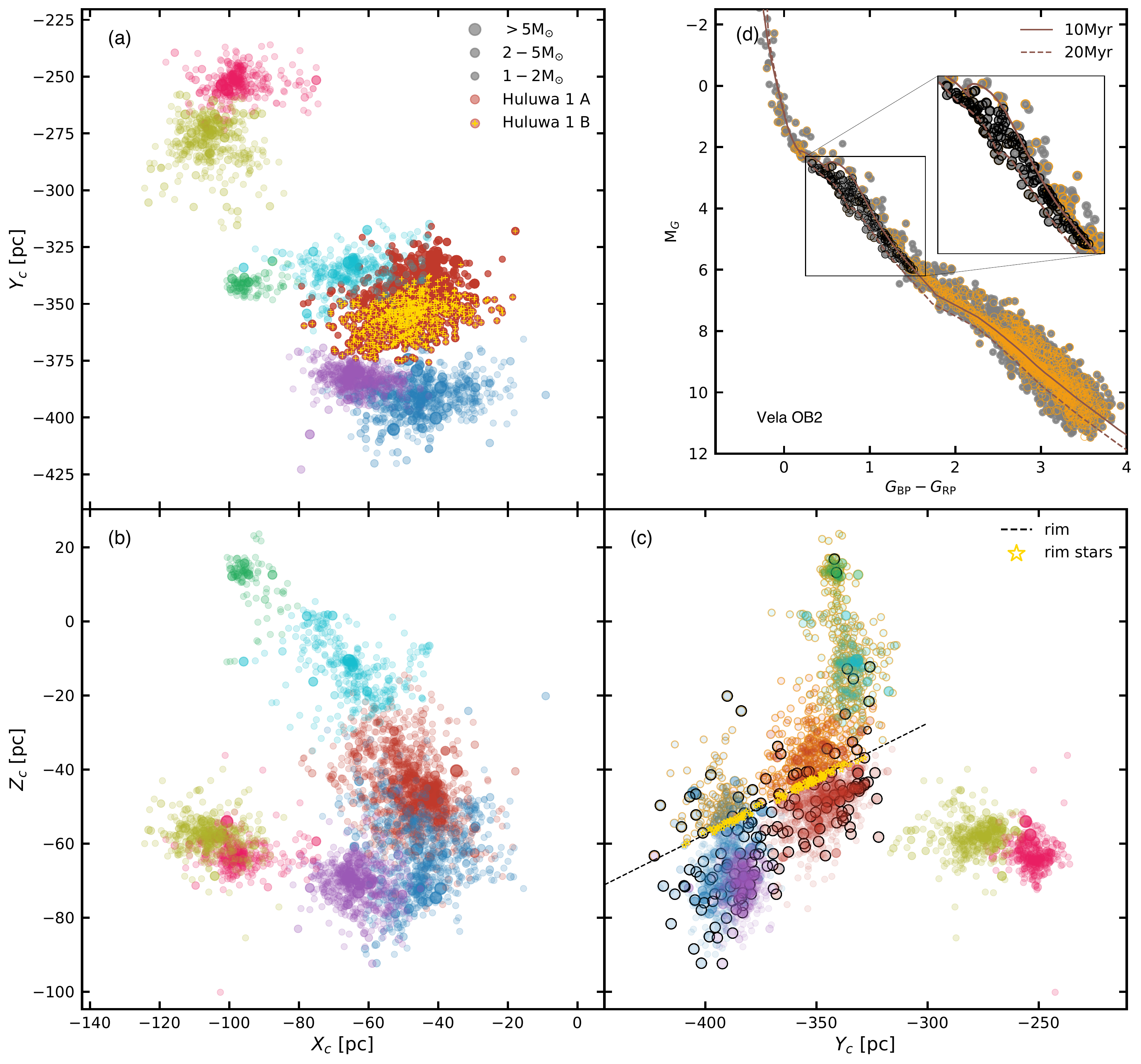}
\caption{(a)--(c): Distribution of 3D spatial position of members in Huluwa\,1--5, Collinder\,135 and UBC\,7 in heliocentric Cartesian coordinates, after distance correction using a Bayesian approach. The color coding is identical to that in Figure~\ref{fig:PM}. We separate Huluwa\,1B as yellow crosses and Huluwa\,1A as red dots in (a). In (b) and (c) the red dots represent Huluwa\,1 instead. The dashed straight line in (c) is the proposed location the rim of the shell, which is used to separate the upper shell and the lower shell. Yellow-star symbols highlight stars with distance within 1\,pc (along $Z$ axis) from the rim of the shell (dashed line).  
 Panel (d) is the color-magnitude diagram for Vela OB2 members (grey dots). The stars located at the upper shell region is highlighted as orange circle. Upper PMS stars with M$_{G}$ between 2.5 and 6\,mag and older (bluer) than the 10\,Myr-old isochrone is indicated as black circles, which are mainly located at the lower shell region (c). The inset is a zoom-in version of the upper PMS region. }
\label{fig:3d}
\end{figure*}

In Figure~\ref{fig:3d}, panels (a) to (c), we show 
the projections of the members of Huluwa\,1--5 in Vela OB2, Collinder\,135 and UBC\,7, onto the $X-Y$, $X-Z$ and $Y-Z$ planes, after distance correction. Among the hierarchical substructures in Vela OB2, Huluwa\,1 is the most massive group (red dots) extending up to 50\,pc in the $X$, $Y$ and $Z$ directions. 
Huluwa 2 (blue dots) has an extent of $\sim$60--70\,pc especially along the $Z$ direction. Huluwa\,3 (purple dots) is more compact than Huluwa\,1 \& 2. Huluwa\,4 (cyan dots) has an elongated shape, connecting the largest cluster, Huluwa\,1, and the smallest cluster, Huluwa\,5  (green dots). 

The cluster pair, Collinder\,135 and UBC\,7 are separated along the $Y$ direction, with separate concentrations of stars.  Collinder\,135 has an extended tail ($\sim50$\,pc long) along the $Y$ axis, while the tail of UBC\,7 is oriented along the $X$ direction, and is 40\,pc in length. The centers of both clusters are separated by only 25\,pc. 

Similar to the cluster pair, the two components of Huluwa\,1, Huluwa\,1A \& 1B, can be divided along $Y$ direction at $Y\sim$--360\,pc. Both Huluwa\,1A (red dots in Figure~\ref{fig:3d}) and Huluwa\,1B (yellow crosses) have a diffuse distribution without any clear central concentration. The centers of Huluwa\,1A and Huluwa\,1B (see Table~\ref{tab:general}) are separated by merely 13\,pc, which is significantly closer than the previous value of 38\,pc reported by \citet{franciosini2018}. Their spatial separation of less than 30\,pc \citep{dela_fuente2009a}, the similarity of their PM and RV distributions, and their almost identical mass, suggest that Huluwa\,1A and Huluwa\,1B may be a coeval cluster pair candidate, resembling the cluster pair Collinder\,135 and UBC\,7. 

\begin{figure*}[tb!]
\centering
\includegraphics[angle=0, width=1.\textwidth]{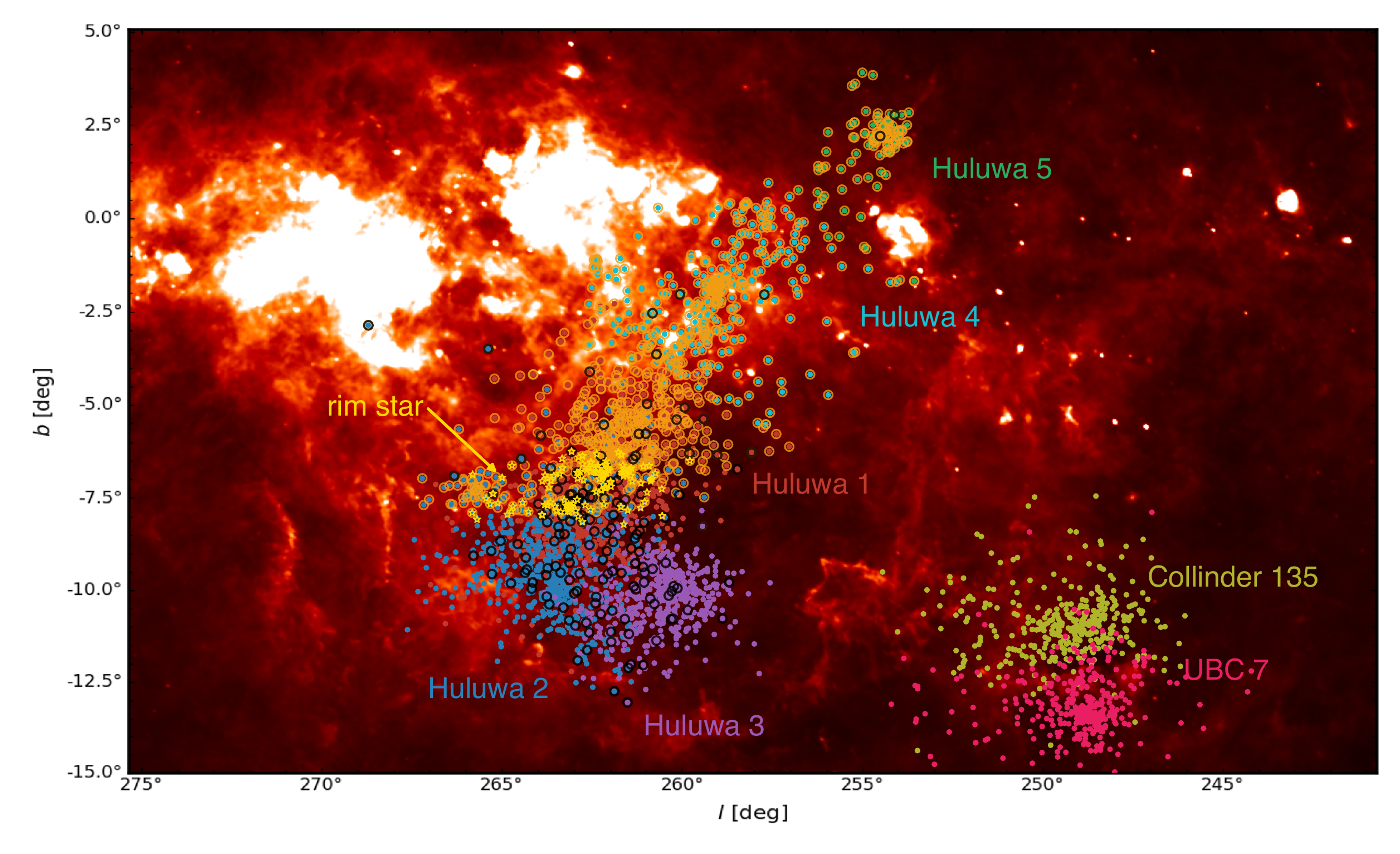}
\caption{IRAS-IRIS infrared image of the 60\,$\rm \micron$ band \citep{miville2005}. The members of Huluwa\,1--5, and the cluster pair Collinder\,135 and UBC\,7 are displayed as colored dots.  Yellow-star symbols highlight stars located along the rim of the shell, within 1\,pc (along $Z$ axis) from the rim (dashed line in Figure~\ref{fig:3d} (c)).  The color coding is identical to that in Figure~\ref{fig:3d}. Stars located in the upper shell region (Figure~\ref{fig:3d}) are outlined with orange circles. The high-mass upper PMS stars older than 10\,Myr shown in the insect of Figure~\ref{fig:3d} (d) are highlighted with black circles.}  
\label{fig:lb}
\end{figure*}

Spatial distribution of the members of Huluwa\,1--5 mimic a ring-like structure, which is located approximately along the rim of the IRAS Vela shell \citep{sahu1992,cha2000}. In Figure~\ref{fig:lb} we display the 2D-positions of the identified members, overlayed on the IRIS image \citep{miville2005}, which shows the gas structures in this region. The ring-like structure is also apparent in 2D projection, along the dense gas structure, which might be at the background. 
\citet{cantat2019a} suggested that star formation in Vela OB2 was triggered by a supernova explosion inside the cavity of the IRAS Vela shell. The supernova might be a member of a 30\,Myr old generation of stars \citep[group III in][]{cantat2019b}. However, the remnant of the hypothetical supernova that created the IRAS Vela Shell has not been detected yet.

In panel (c), we highlight the positions of the higher-mass upper PMS stars with M$_G$ magnitudes between 2.5\,mag to 6\,mag and older (bluer) than the 10\,Myr-old isochrone (black circles in Figure~\ref{fig:3d} (d)) as black circles. 
They are mostly located in the lower part of the shell structure. Following the spatial distribution of these upper PMS stars (black circles), we define the rim of the shell as a straight dashed line (panel (c) in Figure~\ref{fig:3d}), and define the stars in Huluwa\,1--5 located above the dashed line as the upper shell, and those below as lower shell. All stars of the upper shell (the colored dots with orange circles outline in Figure~\ref{fig:lb}) form a clear shell structure in the 2D projection. The rim of the shell (yellow stars in Figure~\ref{fig:lb}), appears to agree with the location of the edge of the gas shell at the background.
  
By plotting stars in the upper shell in the CMD in Figure~\ref{fig:3d}~(d) as orange circles, we can see that they are primarily low-mass PMS. We find a deficiency of higher-mass upper PMS stars with M$_G$ magnitudes between 2.5\,mag to 6\,mag in the upper shell region (zoom-in inset in Figure~\ref{fig:3d}~(d)). 
It seems that most of these older and massive PMS stars are located at the lower shell region (Figure~\ref{fig:3d} (c)). On the contrary, stars at the upper shell are mostly younger and of lower mass.

\subsection{Mass stratification along the shell}\label{sec:mass_stra}

\begin{figure*}[tb!]
\centering
\includegraphics[angle=0, width=0.99\textwidth]{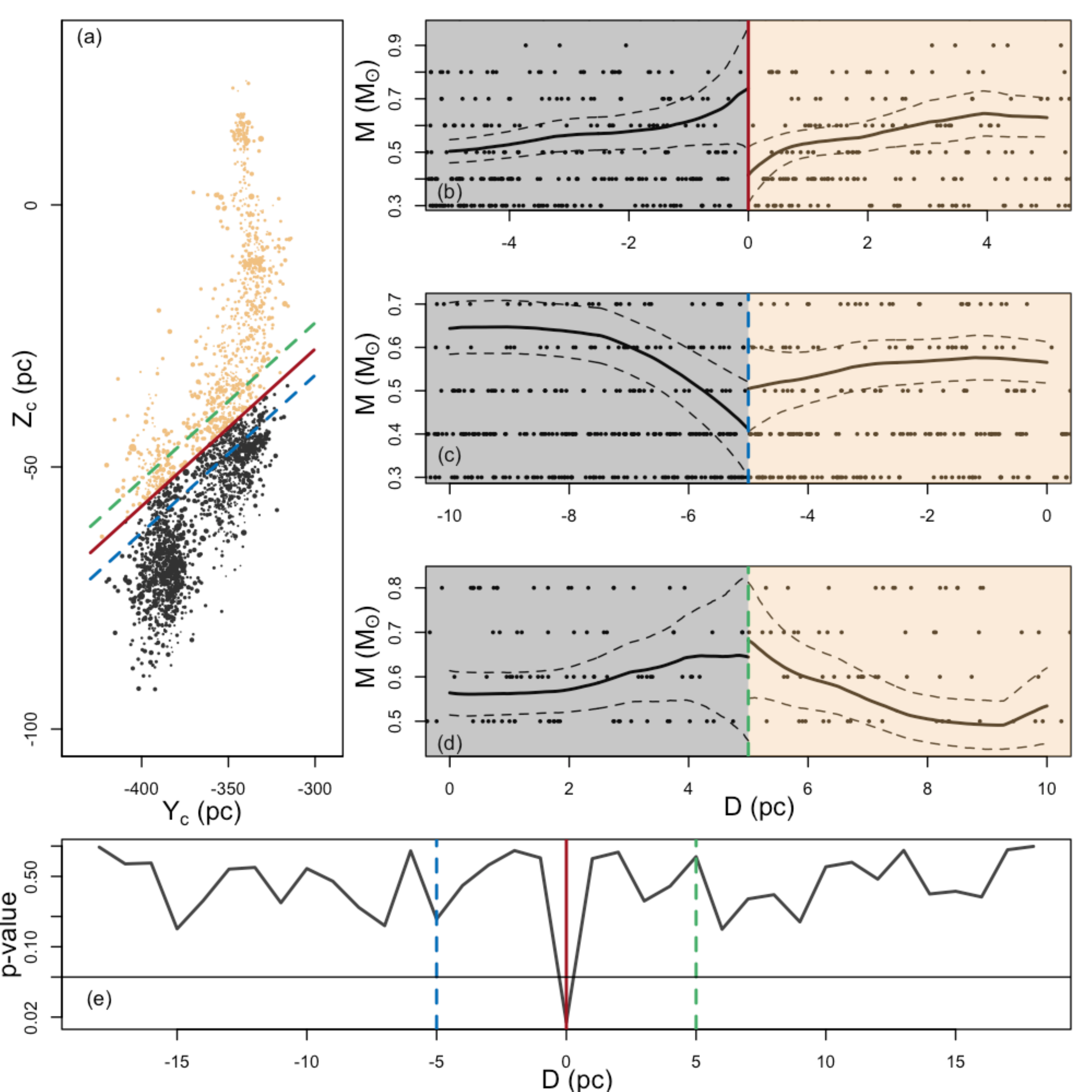}
\caption{
            (a): Distribution of members in Huluwa\,1--5 on the $Y-Z$ plane. The size of each circles is proportional to the mass of a star.
                The  red solid line indicates the rim of the shell in panel~c of Figure~\ref{fig:3d},  which is taken into consideration for 
                the regression discontinuity estimation. Stars in the upper rim of the shell are colored orange and stars in the lower are colored 
                black. The dashed green and blue lines are the rim of the shell (red line) moved upward ($+$5\,pc) and downward ($-$5\,pc) along 
                the $Z$-axis respectively. 
            (b): The mass of stars as a function of the distance of stars within 5\,pc from the rim of the shell (red solid line in (a)). Panels (b) and (d) 
                show the results for the case when the dashed green and blue lines in (a) are taken as the rim positions. The stellar mass is obtained in
                Section~\ref{sec:age} via the $k$-D tree method by taking the mass from the nearest neighbors of the 10\,Myr isochrone. Therefore, the stellar 
                mass is discretely distributed. Negative distance values represent the upper shell (grey) and positive values represent the lower shell (black). The local linear regression curve is obtained from the RD estimation with an associated $95\%$ confidence interval marked by dashed lines. When the confidence areas do not overlap at the rim ($D=0, -5, 5$), it indicates a presence of a discontinuity in the mass distribution on both sides. (e): Dependence of the $p$-value on the distance within 15\,pc from the proposed rim of the shell ($D=0$; red solid line) along $Z$-axis. The $p$-value is computed by moving the rim (the red solid line) along the $Z$-axis in increments of 1\,pc. The green and blue dashed lines indicate  positions of 5\,pc from the rim. The horizontal line indicates  p-value of 5\%. Color coding is identical as panel (a).}
\label{fig:mass_discont}
\end{figure*}

To further quantify mass stratification across the shell, we adopt a regression discontinuity (RD) estimation. The RD method has been used in a number of fields such as medicine and economics, and has only recently been applied in astrophysics \citep[][]{2019RNAAS...3..179P}.

The RD approach is motivated by the need to identify causal effects in observational studies \citep{imbens2008regression}. For example, the causal effect in this case is that the supernova explosion triggers mass stratification across the rim of the shell (red straight line in Figure~\ref{fig:mass_discont} (a)). If the spatial mass distribution is discontinuous at the rim of the shell, this causal effect is very probable. Alternatively, a continuous distribution at the rim means a low probability that the supernova explosion is the cause of observed mass distribution.

We adopt the rim of the shell in Figure~\ref{fig:3d} (dashed line in (c); solid red line Figure~\ref{fig:mass_discont} (a)). Stars located in the upper rim (orange dots in Figure~\ref{fig:3d} (c) and Figure~\ref{fig:mass_discont} (a)) are  located along the Vela IRAS gas shell (orange dots in Figure~\ref{fig:lb}). 
We hypothesize that the masses of stars on either side of the shell should differ only due to the influence of the supernova explosion. We carry out a local linear regression for the mass dependence on the star's distance ($D$) to the rim at the upper ($+D$) side and lower ($-D$) side of the shell (solid curve in (b)--(d) in Figure~\ref{fig:mass_discont}). From comparing the intercepts of the local linear regression on both sides, when the confidence level of the intercepts do not overlap on both sides, we conclude that the causal effect is a likely explanation.

A critical issue in the RD approach is the determination of the optimal bandwidth for the regression, which determines the boundary of the kernel function included in the local linear regression. Small bandwidths may result in a large 
variance of the estimator because few data points are included, while large bandwidths may smooth out the discontinuity. We adopt a prescription to generate the optimal bandwidth by \cite{imbens2012optimal}, which is implemented in the package for the R programming language \citep{rdd}. We obtain a optimal bandwidth corresponding to $2.08$\,pc.

Figure~\ref{fig:mass_discont} (b) displays our results in a graphical form. The stellar mass at the lower side of the rim ($-D$; lower shell) is found to be $0.32 \pm 0.14\,{\rm M_\odot}$ higher than at the upper side ($+D$; upper shell), with an associated $p$-value of $1.8\%$. The $p$-value is a measure for the probability that this finding can be explained by a random occurrence. A $p$-value less of than 5\% is can be interpreted that the proposed causal effect is very likely. 
The estimated difference in the intercepts remains robust, even when we reduce the bandwidth to half ($p = 0.45\%$) or double the value ($p=4.3\%$). When the position of the rim of the shell (solid red line in Figure~\ref{fig:mass_discont} (a)) is varied by moving it upward (green dashed line) or downward (blue dashed line) by 5~pc along $Z$ axis, the resulting $p$-value becomes less significant and the intercepts on both sides of the rim overlap (in Figure~\ref{fig:mass_discont} (c) and (d)). We vary the location of the rim (with steps of 1\,pc) from the proposed position (solid red line in Figure~\ref{fig:mass_discont} (a)) by 15\,pc along the $Z$-direction, and find that the mass discontinuity is most robust at the proposed rim position, where the $p$-value reaches a minimum (Figure~\ref{fig:mass_discont} (e)).

\begin{figure*}[tb!]
\centering
\includegraphics[angle=0, width=1.\textwidth]{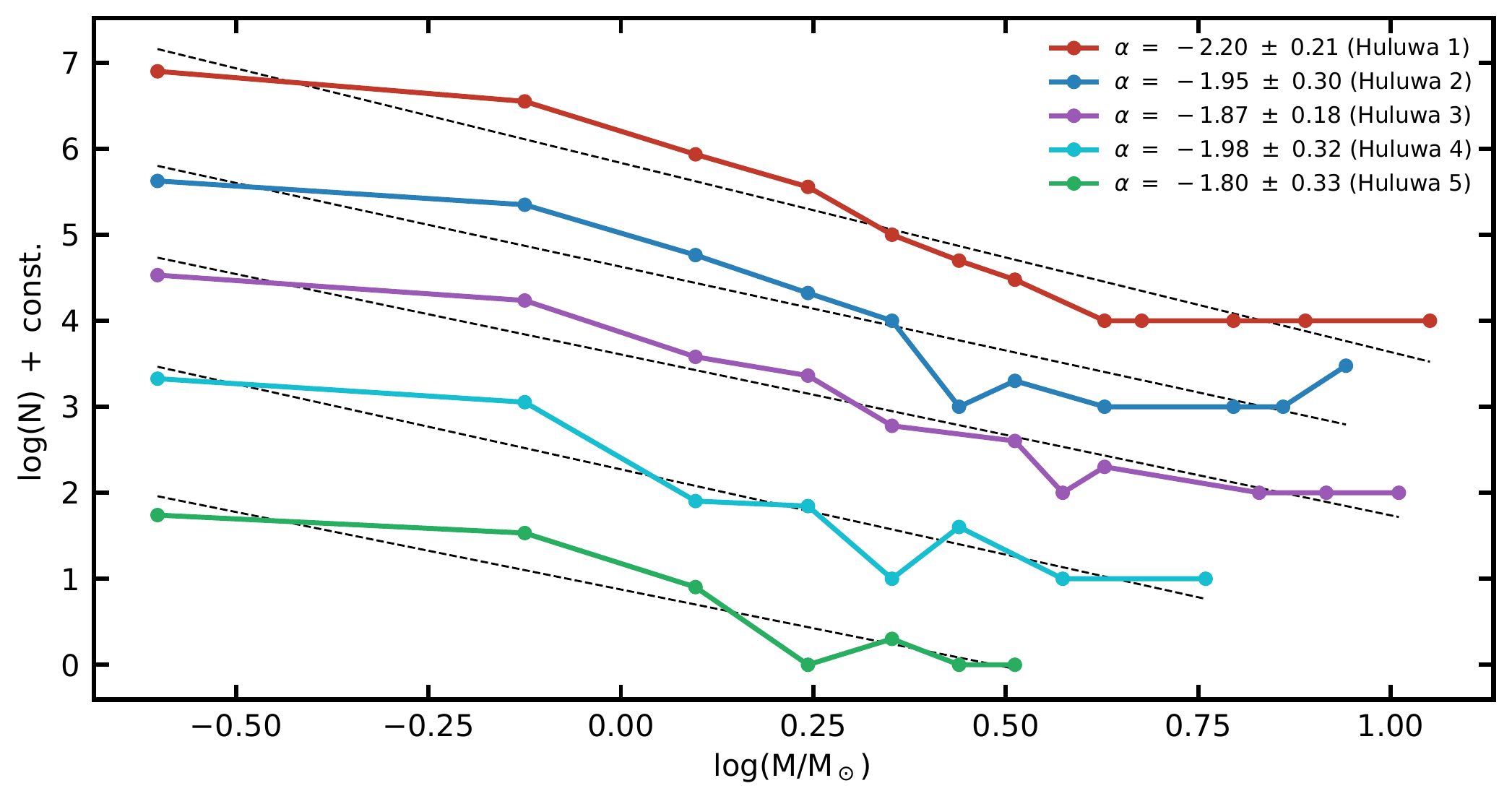}
\caption{The present-day mass function of Huluwa\,1--5, with stellar masses derived from the nearest neighbors of the 10\,Myr isochrone via $k$-D tree method. The Y-axis is shitted manually with a constant so that the mass function of each cluster is visible. The black dashed line is the fitted mass function $dN/dm \approx m^{-\alpha}$ via linear least-squares fitting. The quantity $\alpha$ is the fitted slope, and is indicated for each cluster in the upper-right corner. }
\label{fig:MF}
\end{figure*}

 As Huluwa\,1--3 mainly are located in the lower shell region, while Huluwa\,4--5 are located in the upper shell, we search for evidence in the mass function in each group to cross-check the results obtained using the RD method. We plot the present-day mass function for Huluwa\,1--5 in Figure~\ref{fig:MF}. The slope $\alpha$ in the mass function $dN/dm \approx m^{-\alpha}$ is constrained by using a linear least-squares fit. The slope of the mass function in the five groups are in an agreement when the errors are taken into consideration. There is an apparent deficiency of high-mass stars in Huluwa\,4--5, which is expected from \citet{weidner2013}'s model in which more massive clusters tend to have more high-mass stars. The observed maximum stellar mass is 11\,$\rm M_\odot$ in Huluwa\,1. If an age 20\,Myr (upper limit) is adopted for Huluwa\,1, the maximum stellar mass for a star remains in the MS phrase is 12\,$\rm M_\odot$, while higher-mass stars would have evolved off the MS  \citep{banerjee2020}. 
This is consistent with the age spread  in Huluwa\,1--5. However, if the massive OB stars have been expelled from the cluster \citep{wang2019}, or stars more massive than 20\,$\rm M_\odot$ have directly evolved into black holes \citep{banerjee2020}, the observed 11\,$\rm M_\odot$ does not represent the maximum stellar mass. 

Huluwa\,1--3 contain more stars above 3\,$\rm M_\odot$ than Huluwa\,4--5. This confirms that  more high-mass stars are in the lower shell where Huluwa\, 1--3 are located than in the upper shell that mainly contains Huluwa\,4--5.

\subsection{Mass Segregation}\label{sec:seg}

As high-mass stars are stratified along the rim of the shell (dashed lines in panel (c) in Figure~\ref{fig:3d}), we adopt the Minimum Spanning Tree (MST) method \citep[see][for details]{allison2009} to further quantify the degree of mass distribution in Vela OB2 globally, and in the cluster pair. 3D spatial positions of members \citep{pang2021} are adopted to avoid the dependence of the results on the viewing angle in 2D. 

\begin{figure}[tb!]
\centering
\includegraphics[angle=0, width=1.\columnwidth]{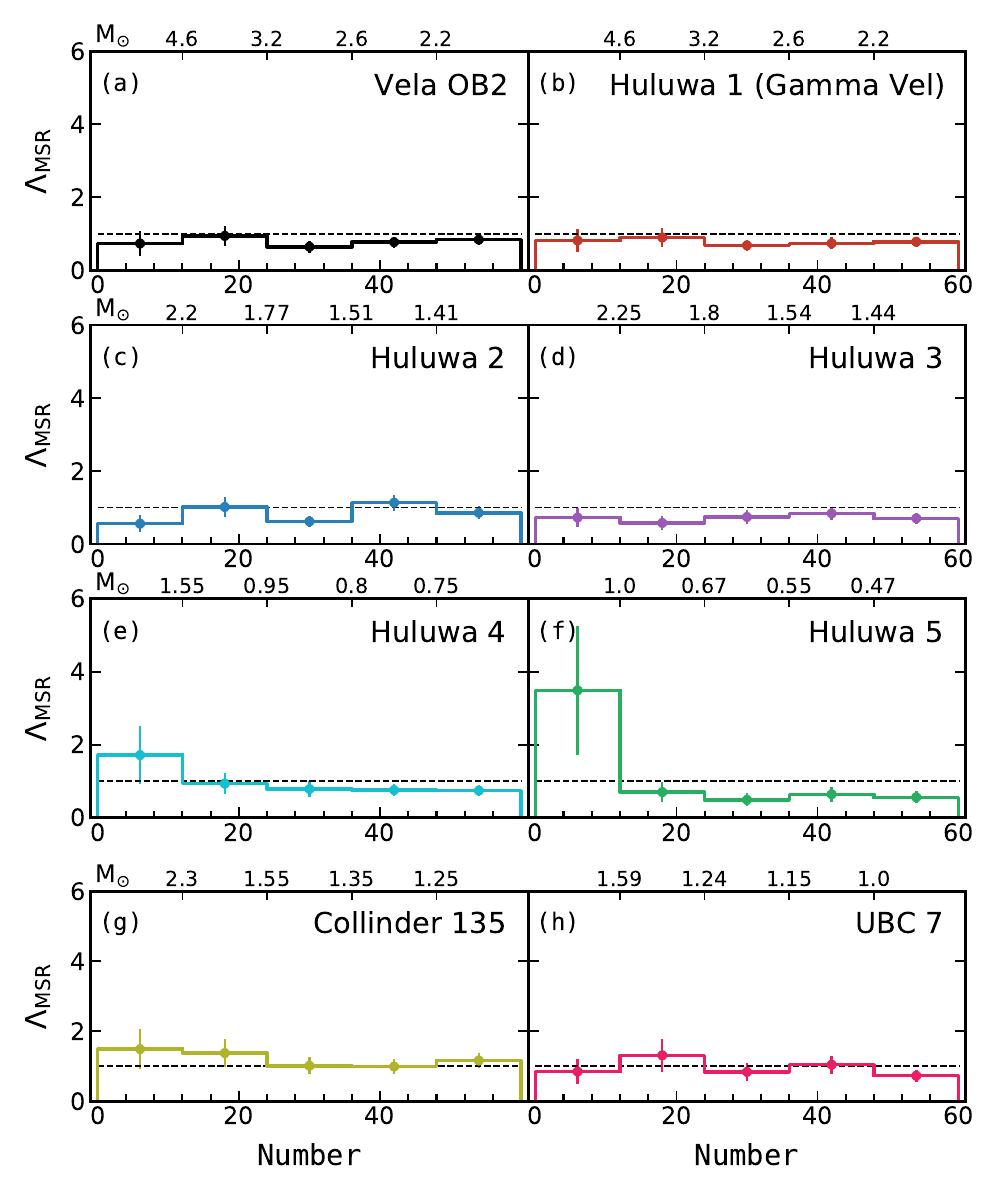}
\caption{The ``mass segregation ratio'' ($\Lambda_{\rm MST}$) for the 60 most 
        massive members, with a bin size of 12 stars in the whole Vela OB2 region (a) for each of the individual clusters Huluwa\,1--5 (b--f), and for the cluster pair Collinder\,135 (g) and UBC\,7 (h).
        The dashed line ($\Lambda_{\rm MST} = 1$) indicates an absence of mass segregation. Increasing value of $\Lambda_{\rm MST}$ indicates a more significant degree of mass segregation. The error bars indicate the uncertainties obtained from a hundred realizations of randomly selected stars.}
\label{fig:mst}
\end{figure}

The MST method compares the minimum path length among the $N$th most massive members of a stellar grouping to that of the minimum path length of $N$ random members. The method is insensitive to massive stars with  extended configuration like a ``ring'', ``cross'' or ``clump'', unless a higher weight is assigned to the dominant configuration \citep{olczak2011}. 
Although massive stars are stratified by the rim of the shell (Section~\ref{sec:mass_stra}) in Vela OB2, such extended configuration cannot be detected using the MST method (black histogram in Figure~\ref{fig:mst}~(a)).

We again apply the MST method to individual group Huluwa\,1--5 (panels (b)--(f) in Figure~\ref{fig:mst}). We find mass segregation down to 1\,M$_{\sun}$ in the most compact but lowest-mass group, Huluwa\,5. The other more massive groups Huluwa\,1--3 show no clear evidence of mass segregation. The degree of mass segregation in Huluwa\,4, Collinder\,135 and UBC\,7 is marginal. 

Mass segregation in a star cluster can be both primordial or dynamical. Dynamical mass segregation occurs as a consequence of the equipartition of energies of stars during two-body encounters. The timescale of dynamical mass segregation is proportional to the relaxation time, which in turn depends on the number of stars, the half-mass radius and the total mass of the cluster. Therefore, the smallest group Huluwa\,5 has the shortest relaxation timescale of $\sim$ 3.8\,Myr, computed using equation~(7.108) in \citet[][adopting the half-mass radius, the total mass, and the number of members from Table~\ref{tab:general}]{binney2008}. The age of 10\,Myr may already allows for a dynamical segregation taking place in Huluwa\,5. Even though primordial segregation may have been present in these groups at early times, the violent binary interactions in the center that expel massive stars, or a significant expansion can altogether erase the observable signatures of primordial mass segregation.

\section{Dynamical State}\label{sec:dyn}
\subsection{Expansion in the Hierarchical Structures}\label{sec:expand}

\begin{figure*}[tb!]
\centering
\includegraphics[angle=0, width= 1. \textwidth]{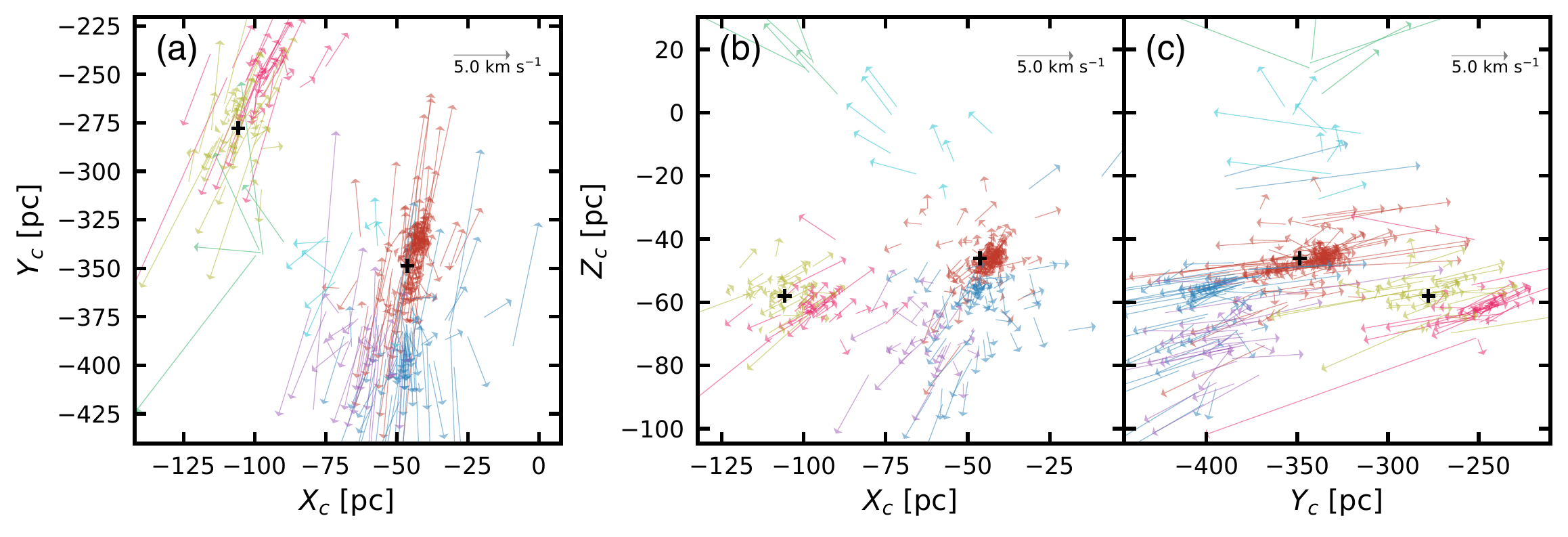}
    \caption{The relative 3D velocity vectors for members, projected onto the $X$-$Y$, $X$-$Z$ and $Y$-$Z$ planes. The red, blue, purple, cyan and green vectors represent the velocities of member stars in Huluwa\,1--5, respectively, relative to the mean motion of Huluwa\,1. The yellow and pink vectors represent the velocities of member stars in Collinder\,135 and UBC\,7, relative to the mean motion of Collinder\,135.
The centers of Huluwa\,1 and Collinder\,135 are indicated with (+) symbols. The scale of the velocity vectors is indicated in the upper-right corner of each panel.  }
\label{fig:vel_vector}
\end{figure*}

\begin{figure*}[tb!]
\centering
\includegraphics[angle=0, width= 0.85\textwidth]{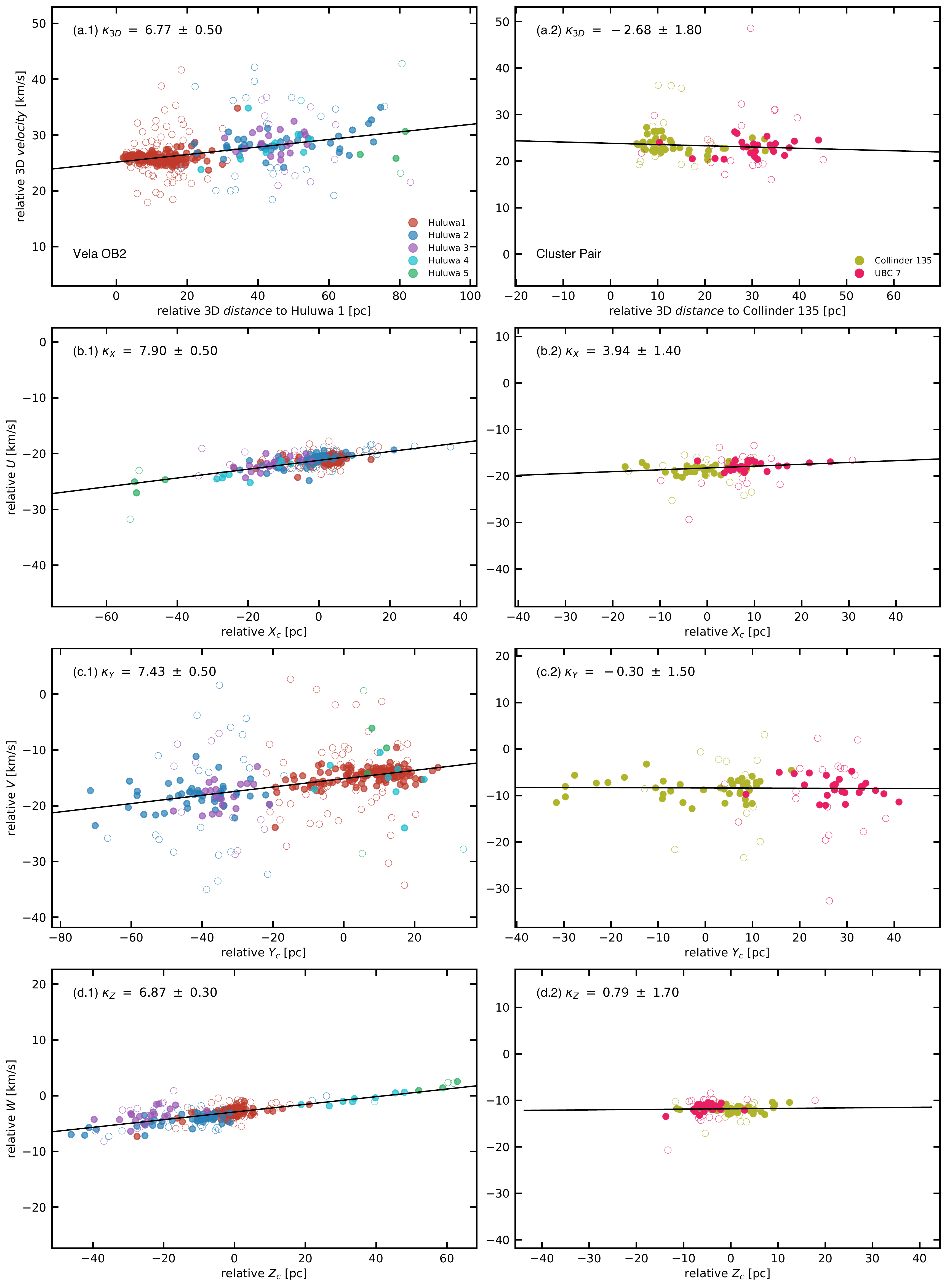}
    \caption{The relative 3D and 1D velocities from Figure~\ref{fig:vel_vector} versus the relative 3D and 1D distances (along the $X$, $Y$ and $Z$ axes) to the center of Huluwa\,1 (left-hand panels) and to the center of Collinder\,135 (right-hand panels). The color coding is identical to that of Figures~\ref{fig:PM} and~\ref{fig:vel_vector}. We fit a linear relation (solid black line) to the members with velocities within the 14 and 86 percentiles in each distance bin (solid colored circles). Stars outside this percentile cut (open circles) are potential binary candidates, or stars with large uncertainties in their velocities. The expansion rate $\kappa$ is the slope of the line, which is indicated in the upper-left corner of each panel, in units of $10^{-2}$\,km\,$\rm s^{-1}\,pc^{-1}$.
      }
\label{fig:exp_rate}
\end{figure*}

\citet{cantat2019b} report significant expansion within the Vela OB complex (group~VII in their paper, corresponding to Huluwa\,1--5 in the current study). 
The expansion rate of Vela OB2 is suggested to be the highest in the Vela-Puppis region \citep{cantat2019b}. Expansion is also observed in the individual clusters that make up Vela OB2; for example, \citet{franciosini2018} and \citet{armstrong2020} detected expansion in population B of Gamma Velorum (Huluwa 1B).

We further investigate the expansion in the 3D velocity space for both Vela OB2 and for the cluster pair. The 3D velocity vectors of member stars are projected onto the $X-Y$, $X-Z$ and $Y-Z$ planes in panels (a)--(c) of Figures~\ref{fig:vel_vector}. We obtain RVs from \citet{jeffries2014} for members in Huluwa\,1 and Huluwa\,2. RVs of the remaining stars are obtained from Gaia DR\,2. The velocity vectors of stars in Huluwa\,1--5 are relative to the mean motion of Huluwa\,1, $(U, V, W)=(-21.1, -14.7, -2.9)$\,km~s$^{-1}$ in heliocentric Cartesian coordinates. In the cluster pair, the velocities are relative to the mean motion of Collinder\,135, $(U, V, W)=(-18.6, -8.6, -12.1)$\,km~s$^{-1}$. A clear sign of expansion is found in Huluwa\,1--5, as the hierarchical clusters are seen to recede from the most massive group (Huluwa\,1). Expansion is also observed in the cluster pair for the first time, with members moving away from Collinder\,135. 

To quantify the expansion, we show the relative 3D velocity ($U, V, W$) of these stars versus the relative distance to the center of Huluwa\,1 (left-hand panels) or to the center of Collinder\,135 (right-hand panels) along the ($X, Y, Z$) axes in Figure~\ref{fig:exp_rate}. A positive/negative correlation indicates expansion/contraction, according to the linear expansion model by \citet{wright2018}. Unlike previous studies that use all identified members to probe the signature of expansion, here in this study we select stars with 3D velocity value between 14 and 86 percentile in each distance bin (solid colored circles) to fit a linear correlation, so as to avoid the influence of binary candidates. 
We observe a positive relationship between relative velocities and positions, with a 1D expansion rate $\kappa$ (i.e., the slope of the fitted line) ranging from $(6.9\pm0.3)\times10^{-2}$~km\,$\rm s^{-1}\,pc^{-1}$ to $(7.9\pm0.5)\times10^{-2}$~km\,$\rm s^{-1}\,pc^{-1}$ in Vela OB2, which is largest along the $X$-axis and smallest along the $Z$-axis.  This value is more significant than the expansion rate in the 10\,Myr-old Scorpius-Centaurus OB2 region \citep[on average $(2.0-3.5)\times10^{-2}$~km\,$\rm s^{-1}\,pc^{-1}$,][]{wright2018} and in the 6--9\,Myr-old Orion complex \citep[on average $(1.0-1.3)\times10^{-2}$~km\,$\rm s^{-1}\,pc^{-1}$,][]{swiggum2021}. We find higher expansion rates than those obtained by \citet{cantat2019a}, but the findings are  consistent within their uncertainties.  
The 1D expansion rate in the cluster pair is much smaller. The 3D expansion rate even indicates contraction, with a negative value for $\kappa$. The expansion is most pronounced along the $X$-direction, with $\kappa_X=(3.94\pm1.40)\times10^{-2}$~km\,$\rm s^{-1}\,pc^{-1}$. Such anisotropic expansion in Vela OB2 and the cluster pair might be a result of spatially varied stellar feedback \citep{fujii2021b,fujii2021a}. 

Compared to the expansion in a young cluster with similar age \citep[NGC\,2232; see][]{pang2020}, the expansion in Vela OB2 is more significant, and is comparable to that of disrupting clusters with tidal tails \citep[e.g., Coma Berenices,][]{tang2019,pang2021}. 
 The degree of expansion experienced by a cluster does not only depend on how fast the natal gas is dispersed (impulsive versus adiabatic) but also on the SFE \citep{baumgardt2007, dinnbier2020a, dinnbier2020b}. A SFE of 
at least 33\% is required for the remaining cluster to reach equilibrium again after an impulsive gas expulsion event has occurred \citep{baumgardt2007}. 
The high expansion rate in Vela OB2 may indicate not only fast gas removal, but also a SFE below  $\sim 33\%$. 

\subsection{Velocity Dispersions}\label{sec:disp}

To determine whether these hierarchical structures will remain gravitationally bound after such pronounced expansion,  
we compute the dispersion of RVs and PMs of each cluster, to quantify their dynamical states. We follow the procedure  described in \citet{pang2021}, assuming that the likelihood function for the RV distribution can be modeled with two Gaussian components \citep[cluster members and field stars; equations~1 and~8 in][]{cottaar2012}, with the effects of binary broadening and measurement uncertainties taken into account. The PM distribution follows a very similar likelihood function \citep[equations 1--3 in][]{pang2018} as the RV distribution, except for binary correction. For Huluwa\,1--2 (including Huluwa\,1A\&1B), we only use RVs from GES in the computation, while for the other groups we use Gaia DR\,2. Due to the relatively small number of RV measurements in Huluwa\,4 and 5, we do not calculate RV dispersion for these two clusters. We apply the Markov Chain Monte Carlo (MCMC) method to obtain the best-fit values and the corresponding uncertainties. The final dispersions of the PMs and RVs are presented in Table~\ref{tab:general}. 
 
The dispersion of the RV distribution tends to be larger than that of the PMs. Though the global expansion after gas expulsion may induce a velocity anisotropy \citep{baumgardt2007} in Huluwa\,1--3 and the cluster pair, this different dispersions more likely have originated from smaller uncertainties in PM measurements with respect to those of the RVs. Higher-resolution spectroscopy is required to justify this trend. 

Based on the velocity dispersion, we estimate the virial ratio  (the ratio between the total kinetic energy and the total gravitational energy) for Huluwa\,1--3, and for the cluster pair. The resulting virial ratio is larger than 10 (where a value of 0.5 corresponds to a virialized system). This requires at least an additional 20 times more mass to retain the expanding star. Even observational incompleteness cannot conceal such a large amount of mass. Incompleteness often accounts for a few percent of the mass of the cluster \citep{tang2019}. 
This is not surprising, since many OB associations are known to be supervirial, with virial ratios ranging from 10 to 1000 \citep{melnik2017,wright2020}.
Over half of the groups have half-mass radii larger than their tidal radii \citep[Table~\ref{tab:general}, equation 12 in][]{pinfield1998}, which further confirms their disrupting states.

We also measure the velocity dispersion for the two components that make up Huluwa\,1 (i.e., Huluwa\,1A\&1B). Huluwa\,1B has a velocity dispersion that is almost three times higher than that of Huluwa\,1A. The virial ratio is 3 for Huluwa\,1A, and 9 for Huluwa\,1B. The dispersion error causes 10\% uncertainty in the virial ratio. Having 50\% of its mass already unbound (the half-mass radius is identical to the tidal radius; see Table~\ref{tab:general}), Huluwa\,1A appears to be in an early stage of disruption, while Huluwa\,1B (the half-mass radius is 30\% larger than the tidal radius) is already in an advanced stage of disruption. 
 Unlike in previous studies \citep{jeffries2014,franciosini2018,armstrong2020}, we find that Huluwa\,1A may not be a bound core. The entire cluster Huluwa\,1 is supervirial, although Huluwa\,1A is probably dispersing at a slower rate than Huluwa\,1B. The velocity dispersion and virial ratio confirm the dispersing state of the Vela OB2 structures and the cluster pair, which may be caused by impulsive gas expulsion with a SFE that is well below 33\% \citep{baumgardt2007,dinnbier2020a,dinnbier2020b}.

\section{Star formation history in Vela OB2: the supernova quenching scenario}\label{sec:formation}

Hierarchical structures can be formed in complexes through two channels. The first channel is mild and progressive, and occurs via the propagation of star formation  \citep{elmegreen2000,kerr2021}. The stellar feedback of the first generation of stars excited turbulence in the ambient molecular cloud and produces density fluctuations that drives the cloud to collapse and form stars \citep{maclow2004,fujii2021a}. This turbulence driven star formation often occurs along filamentary structures. The second formation channel, on the other hand, is violent, with supernova shocks compressing gas and generating overdensities. Star formation following this mode often takes place along the rim of a shell around the supernova \citep{kounkel2020}.

We propose an alternative scenario in which the age and spatial configuration of the stars in Vela OB2 (Huluwa\,1--5) is a consequence of sequential star formation.
The supernova inside the Vela IRAS shell \citep{cha2000} plays a destructive role in quenching star formation \citep{krause2016,wang2020}. 
The formation of the 20-Myr old generation in Huluwa\,1--3 may have been triggered by the interplay of stellar feedback from surrounding older clusters that powered HII regions by pushing gas to the periphery \citep{fujii2021b}, such as Collinder\,135/UBC\,7.  
 
 After about 3\,Myr, the massive stars in Huluwa\,1--3 progressively produced stellar feedback and pushed the gas away from the central clusters 
 and excited turbulence in the molecular cloud. This generated local density fluctuations along the filamentary structures \citep{maclow2004,fujii2021b} and produced Huluwa\,4--5. Simultaneously, the stellar population of Huluwa\,1--3 started to expand after the gas expulsion phase; the majority of their members are now located in the lower region of the shell. When Huluwa\,4--5 were forming, explosion of a supernova (may be a member of a 30\,Myr old generation of stars \citep[population III in][]{cantat2019b}) quenched the star formation by efficiently exhausting the gas in the upper shell region, which is primarily the location of Huluwa\,4--5. Therefore, most low-mass proto-stars in this region stopped accreting gas, and failed to grow into massive stars, which require a longer accretion time than their lower-mass counterparts \citep{walker2021}. This violent removal of gas has substantially accelerated the disruption process of Huluwa\,1--5. 

\begin{figure*}[p!]
\centering
\includegraphics[angle=0, width= 1.\textwidth]{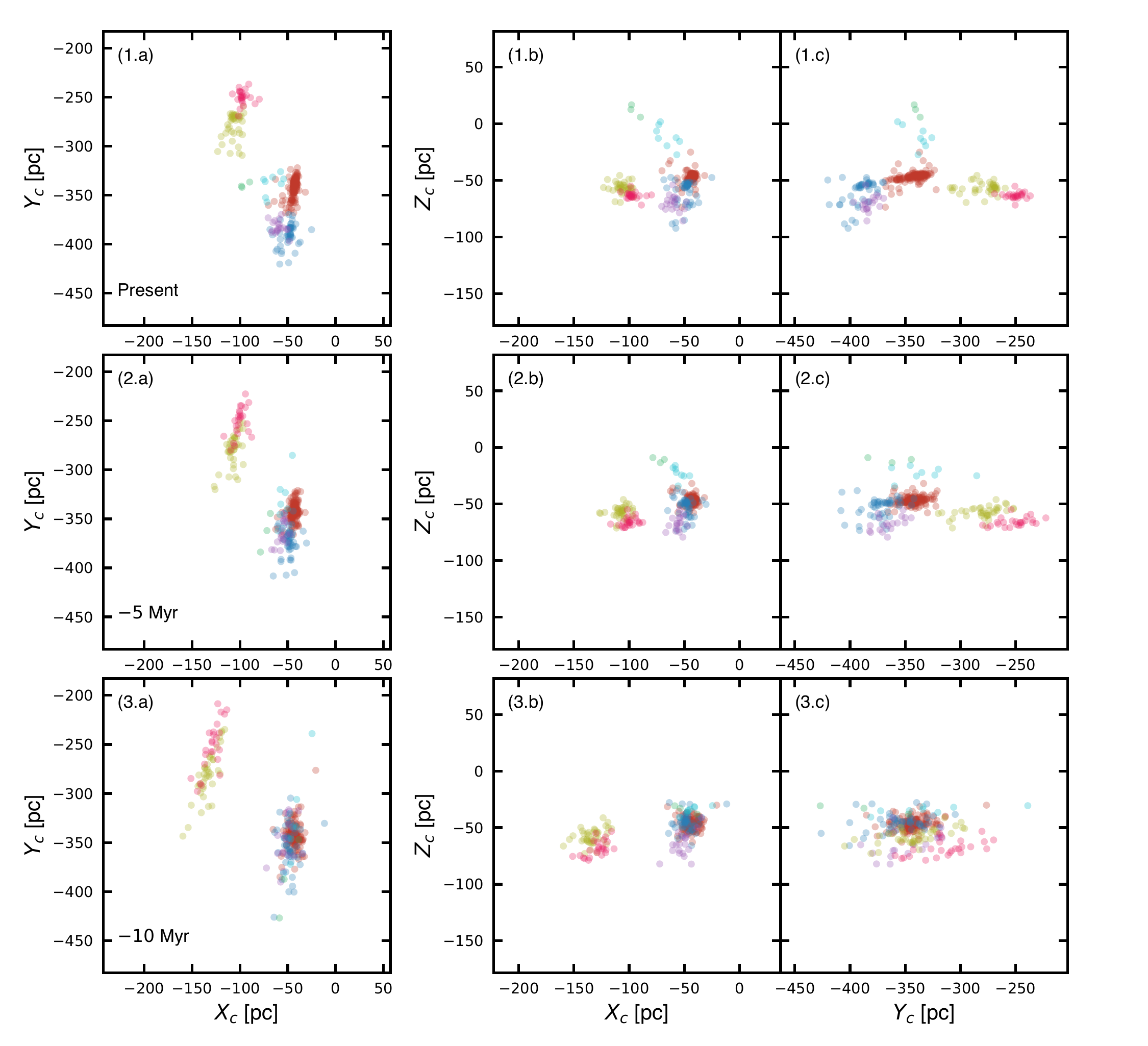}
    \caption{3D configuration of Vela OB2 and the cluster pair, from the present to past 10\,Myr.  To trace back --5\,Myr and --10\,Myr, we only use members with velocities within the 14 and 86 percentiles in each distance bin (solid colored circles in Figure~\ref{fig:exp_rate}), and assume that stars follow linear motions with their present-day observed 3D velocities. The panels in the first row show the current positions of these members. The color coding is identical to that of  Figure~\ref{fig:exp_rate}.
      }
\label{fig:date_back}
\end{figure*}

To further investigate the sequential star formation and supernova quenching scenario, we date back the position of the member stars based on their current positions and 3D velocities  \citep[see Figure~\ref{fig:date_back}, assuming a linear motion for members,][]{wright2018}. Only the solid colored circles in Figure~\ref{fig:exp_rate} are used in this calculation. If Huluwa\,1--5 were formed from gas compressed by the expanding shell due to the shock front of the hypothetical supernova, all stars should return to the central region of the cavity inside the Vela IRAS shell \citep[][]{kounkel2020}. On the contrary, if stars of Huluwa\,4--5 were formed from turbulence-driven over-densities, they would follow the filaments and return to the source that empowered the turbulence (Huluwa\,1--3). 
As can be seen from Figure~\ref{fig:date_back}, the filament-shape-like Huluwa\,4--5 shrunk back to a small space close to Huluwa\,1 in the recent 10\,Myr, like other  two clusters Huluwa\,2--3.
This evidence supports the hypothesis that sequential star formation occurred in Vela OB2. The stellar feedback of the more massive group Huluwa\,1--3 triggered turbulence and formed the less massive clusters Huluwa\,4--5.

However, considering the uncertainty in both the position and the distance of the undetected supernova remnant, the proposed scenario provides an alternative formation channel for Vela OB2. We cannot exclude the scenario of supernova triggering star formation  \citep{cantat2019a}, or non-uniform gas density based on the current analysis. Further observational detection of the supernova and observational constraints on its distance and position will certainly aid in identifying the formation history of the stellar structures in this region.

\begin{figure*}[tbh!]
\centering
\includegraphics[angle=0, width=1.\textwidth]{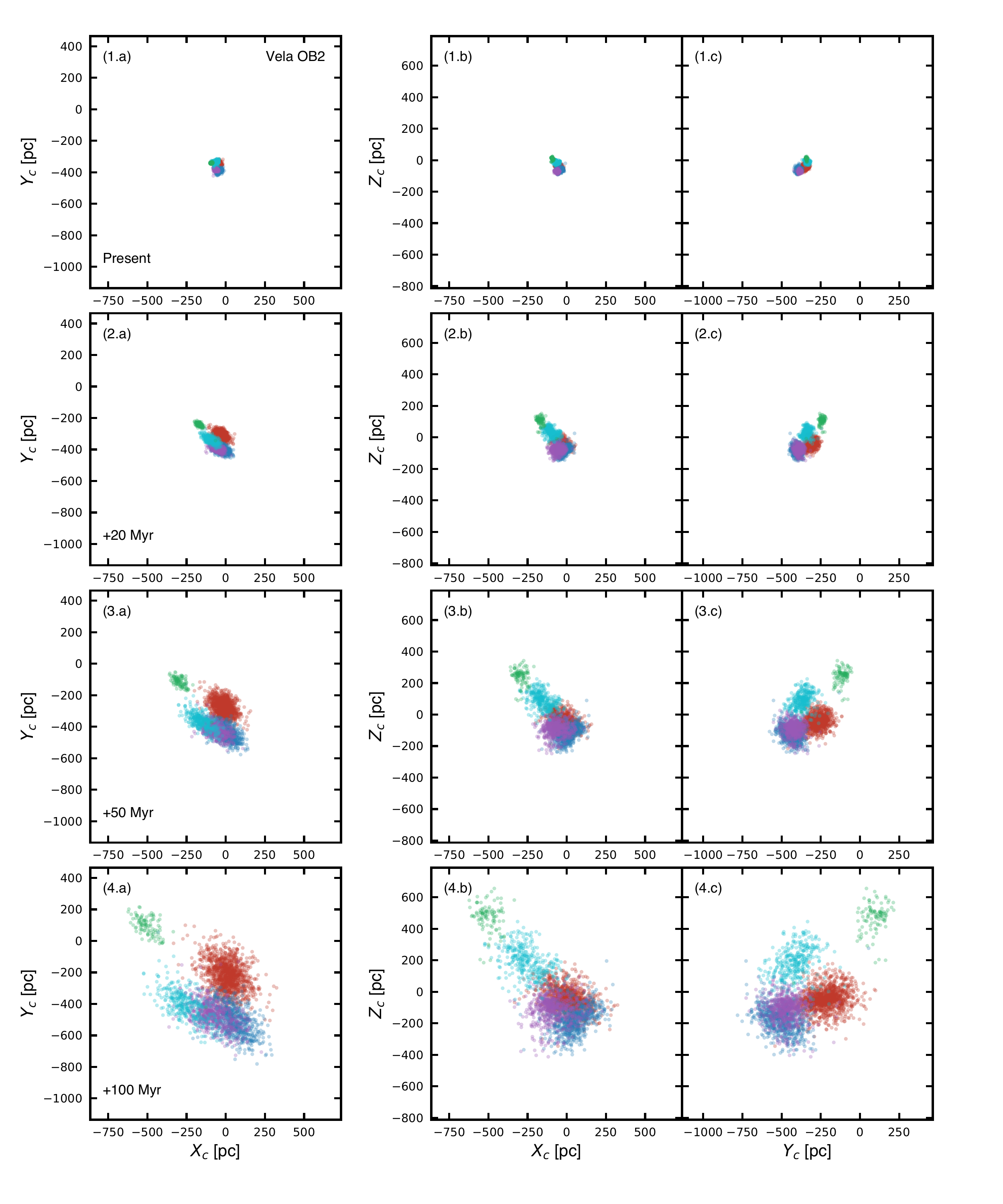}
\caption{3D spatial positions for member stars in the Vela OB2 complex (Huluwa\,1--5) from $N$-body simulations, which start from the present-day positions (panels 1) and evolve into future for 20\,Myr (panels 2), 50\,Myr (panels 3) and 100\,Myr (panels 4). Here, we present the model in which the RV dispersions are inferred from those of the PMs, under the assumption of an isotropic velocity distribution. Vela OB2 continues to expand, and its components show no signs of mutual gravitational interactions. The color coding is identical to that of Figure~\ref{fig:PM}. }
\label{fig:sim_vela}
\end{figure*}

\begin{figure*}[tbh!]
\centering
\includegraphics[angle=0, width=1.\textwidth]{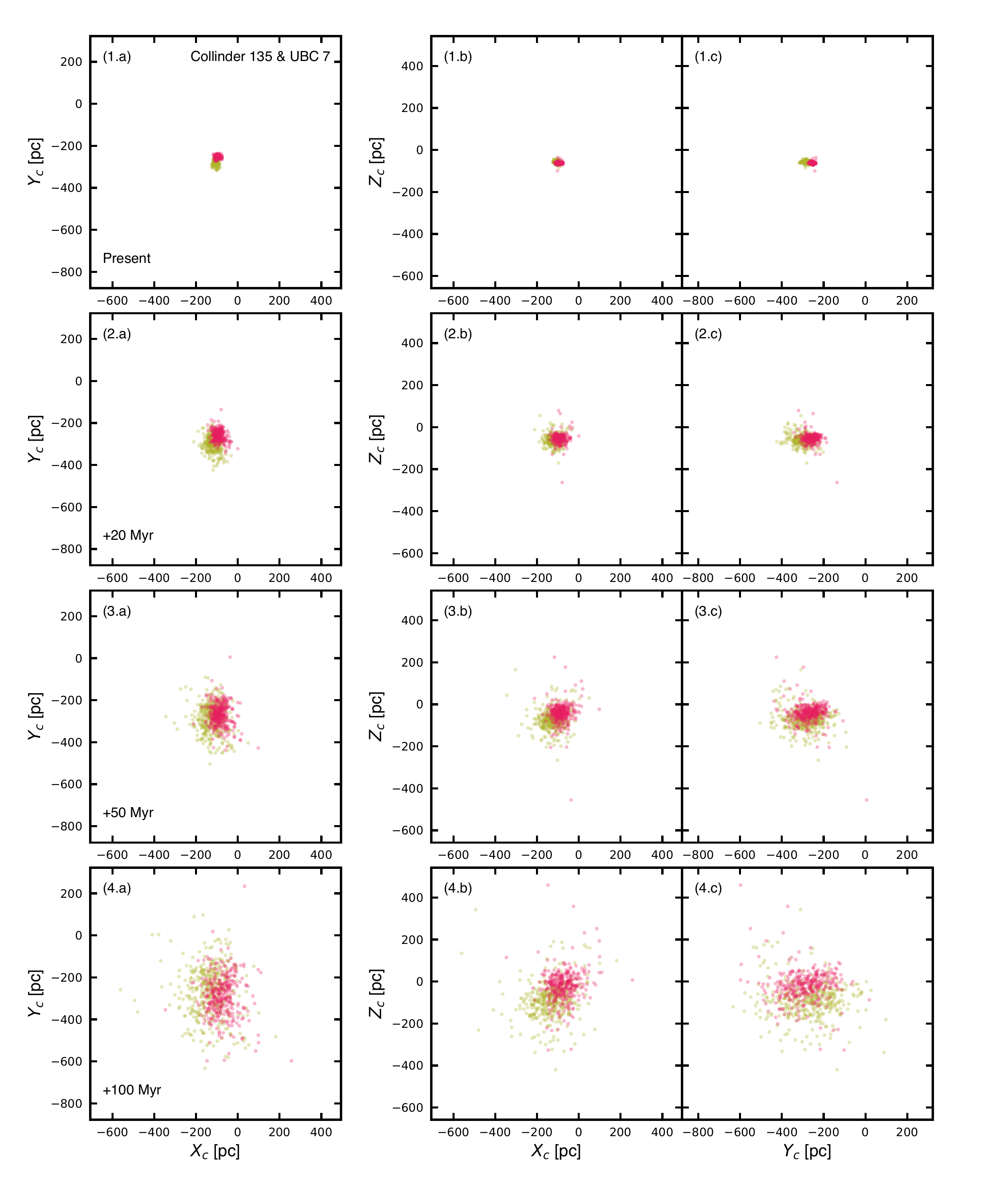}
\caption{3D spatial position for member stars in the cluster pair Collinder\,135 and UBC\,7 from $N$-body simulations, which start from the present positions (panels 1) and evolve into future 20\,Myr (panels 2), 50\,Myr (panels 3) and 100\,Myr (panels 4). Here we present the model in which the dispersion of the RVs is obtained from that of the PMs under the assumption of isotropy. Collinder\,135 and UBC\,7 continue to expand and show no signs of merging or interaction. The color coding is identical to that of Figure~\ref{fig:PM}. }
\label{fig:sim_ubc}
\end{figure*}

\section{Future predictions using $N$-body simulations}\label{sec:nbody}

In order to investigate the dynamical fates of Vela OB2 and the cluster pair Collinder\,135 and UBC\,7, we perform $N$-body simulations using the NBODY6++GPU code \citep{wang2015,wang2016}. We aim to investigate the mutual interaction among clusters Huluwa\,1--5, and between Collinder\,135 and UBC\,7, and determine whether they will eventually merge to become one large structure due to gravitational interactions between sub-clusters as suggested in literature \citep{goodwin2004,allison2009, smith2011,farias2018}. We use the observed members of each group with their 3D spatial configuration after distance correction (Section~\ref{sec:space}) as the initial positions. To assign each star with individual 3D velocity, we make use of PMs and RVs from Gaia EDR3, and RVs from GES. However, only a small fraction of members have RV measurements. 

To correct for the lack of RVs, we assume that the RV distributions of members in each group are Gaussian, with a mean value obtained from the observed average value and the dispersion based on PMs from Gaia EDR3 as a reference value. We evaluate models with three different RV dispersions. First, we study the case in which the RV dispersion is identical to that of the PMs, under the assumption of isotropic motion. We subsequently study models with a velocity dispersion that is half of, or twice, the isotropic value derived from the PM distribution. The RV of each star is randomly assigned for each of the three adopted RV distributions. In our simulations we ignore stellar evolution, stellar binaries, and the tidal effect from the Milky Way that will accelerate the disruption process of the clusters. We also assume that all stellar groupings have been cleared of their gas contents; we therefore do not include any remaining gas in the simulations. 

The simulations are run from the present time into future 100\,Myr. Although the Galactic tide is absent, both Huluwa\,1--5 and the cluster pair continue to expand and end up in a much more diffuse and eventually unbound structure, irrespective of the choice of adopted RV dispersion (Figures~\ref{fig:sim_vela} and~\ref{fig:sim_ubc}). We find that the relative velocities among Huluwa\,1--5, or between Collinder\,135 and UBC\,7, are higher than their internal velocity dispersions, so that there is no chance for them to form a merged system after gravitational interactions \citep{gavagnin2016}. Such significant expansion also prevent any groups in Huluwa\,1--5 (including Huluwa\,1A and Huluwa\,1B) from interacting to become a physical pair. 

According to \citet{kovaleva2020}'s simulations, the physical separation between Collinder\,135 and UBC\,7 was closer in the past than in the present. Our $N$-body simulations show that Collinder\,135 and UBC\,7 will continue to expand, and that their separation will increase in the future 100\,Myr. This cluster pair will eventually become dynamically unbound. The same fate seems to happen to the young cluster pair candidate Huluwa\,1A and Huluwa\,1B. Considering the coevality, the spatial and kinematic proximity between Collinder\,135 and UBC\,7, Huluwa\,1A and Huluwa\,1B, they were most likely formed from the same molecular cloud, i.e., via the fission of a giant molecular cloud induced by oblique cloud-cloud collisions \citep{fujimoto1997}, but not through tidal capture \citep{vandenbergh1996} or resonance of Galactic structures \citep{dehnen1998,dela_fuente2009a}. The latter processes generally result in cluster pairs of different ages. 

Therefore, Huluwa\,1--5 in Vela OB2 and the cluster pair Collinder\,135 and UBC\,7 are all undergoing disruption. Such fast disruption within 10-20\,Myrs  may imply a low SFE (less than 33\%) especially in the region of Vela OB2. 
After the gas evacuation, that was facilitated by the supernova explosion, the shallow potential of the cluster provides no chance to maintain a bound cluster core.

\section{Summary}\label{sec:summary}

\begin{enumerate}

\item We have identified hierarchical structures in the Vela OB2 complex and the star cluster pair consisting of Collinder\,135 and UBC\,7, using data from Gaia EDR3, via the machine learning algorithm \texttt{StarGO}. Five second-level clusters of stars are kinematically disentangled in Vela OB2 (Huluwa\,1, Huluwa\,2, Huluwa\,3, Huluwa\,4, and Huluwa\,5). Huluwa\,1 is the Gamma Vel (or Pozzo\,1) cluster from literature. In total, 3074 member stars of the Vela OB2 complex are identified. 
The clusters Huluwa\,1--3 (10--22\,Myr) are generally older than Huluwa\,4--5 (7--20\,Myr). Collinder\,135 and UBC\,7 are of the same age, 40\,Myr. 

\item The five clusters in Vela OB2 have distinct but coherent proper motion and radial velocity distributions. Consistent with previous studies, Huluwa\,1 is found to consist of two components: Huluwa\,1A and Huluwa\,1B. However, unlike suggestions in earlier works, Huluwa\,1A is not a bound core. Both Huluwa\,1A and Huluwa\,1B are dispersing structures. We suggest that Huluwa\,1A and Huluwa\,1B may be the components of a coeval dispersing binary cluster.

\item Collinder\,135 and UBC\,7 have adjacent proper motion distributions and nearly identical radial velocity distributions.  This confirms previous suggestions that both clusters in this cluster pair have fragmented from the same molecular cloud. 

\item The 3D morphology of Huluwa\,1--5 resembles a shell-like structure, which is elongated along the $Z$-axis, up to 100\,pc. Stars at the upper part of the IRAS Vela shell are younger than those located at the lower shell. A mass stratification is observed and quantified via the regression discontinuity method: more massive stars tend to be located in the lower shell, while low-mass stars tend to be more abundant in the upper shell. The stellar masses is $0.32\pm0.14$\,$\rm M_\odot$ higher in the lower shell region than in the upper shell region.

\item Mass segregation is detected only in the lowest-mass group Huluwa\,5. This mass segregation is likely dynamical evolution, i.e., a consequence of two-body relaxation. Marginal mass segregation is found in Huluwa\,4, Collinder\,135 and UBC\,7. 

\item Significant expansion is observed in the five clusters Huluwa\,1--5 in Vela OB2, with 1D expansion rates ranging from $(6.9\pm0.3)\times10^{-2}$~km\,$\rm s^{-1}\,pc^{-1}$ to $(7.9\pm0.5)\times10^{-2}$~km\,$\rm s^{-1}\,pc^{-1}$. Expansion in the cluster pair is moderate. 1D contraction in the cluster pair is observed along the $Y$-direction.  

\item The velocity dispersions of Huluwa\,1--5, and the cluster pair Collinder\,135 and UBC\,7 are consistent with a state of disruption.  Vela OB2 was probably formed with a star formation efficiency that was less than 33\%. After a phase of violent gas expulsion, the shallow potential is unable to retain the member stars in a gravitationally-bound cluster. 

\item We suggest that the formation of the Vela OB2 complex occurred through a process of sequential star formation, with a younger generation of clusters (Huluwa\,4--5) formed from turbulence triggered by the older generation of clusters (Huluwa\,1--3). The supernova in the cavity of the IRAS Vela shell quenched star formation in the young generation (Huluwa\,4--5) along the upper part of the shell by sweeping away their gas content. This rapid phase of gas expulsion further contributed to the significant expansion observed in Huluwa\,1--5. However, due to the uncertainty in the position of the supernova, the supernova triggering star formation scenario cannot be excluded.

\item $N$-body simulations are carried out to predict the dynamical future of the Vela OB2 structure and the cluster pair. All clusters Huluwa\,1-5 in Vela OB2, and the cluster pair (Collinder\,135 and UBC\,7) will expand for the future 100\,Myr and will eventually dissolve into the Galactic field. 
None of the clusters in Vela OB2 will experience mutual interaction with the other components. Although the clusters Collinder\,135 and UBC\,7, (possibly) Huluwa\,1A and Huluwa\,1B are currently a pair, they will not experience a merger event in the future. 

\end{enumerate}

\acknowledgments
We wish to express our gratitude to the anonymous referee for providing comments and suggestions that helped to improve the quality of this paper. 
X.Y.P. is grateful to the financial support of the research development fund of Xi'an Jiaotong-Liverpool University (RDF-18--02--32); gave thanks to the National Natural Science Foundation of China with grant No: 11503015 and the Natural Science Foundation of Jiangsu Province, No: BK20200252. X.Y.P. also acknowledge the science research grants from the China Manned Space Project with NO. CMS-CSST-2021-A08.
M.B.N.K. expresses gratitude to the National Natural Science Foundation of China (grant No. 11573004) and the Research Development Fund (grant RDF-SP-93) of Xi'an Jiaotong-Liverpool University (XJTLU). 
Z.Y. acknowledge funding from the Agence Nationale de la Recherche (ANR project ANR-18-CE31-0006, ANR-18-CE31-0017 and ANR-19-CE31-0017), from CNRS/INSU through the Programme National Galaxies et Cosmologie. 
M.P.'s contribution to this material is based upon work supported by Tamkeen under the NYU Abu Dhabi Research Institute grant CAP$^3$.

We express gratitude to Prof. Dr. Pavel Kroupa (University of Bonn) for helpful discussion on the gas expulsion, and Prof. Dr. Simon Goodwin (University of Sheffield) for in depth communication on the binary cluster evolution. We give thanks to Dr. Long Wang for discussion on the stellar evolution model and Dr. Chien-Cheng Lin for IRIS data download. We thank Prof. Dr. Youzhou Zhou (Xi'an Jiaotong-Liverpool University) for suggestion on 3D morphology interpretation. 
This work made use of data from the European Space Agency (ESA) mission {\it Gaia} 
(\url{https://www.cosmos.esa.int/gaia}), processed by the {\it Gaia} Data Processing 
and Analysis Consortium (DPAC, \url{https://www.cosmos.esa.int/web/gaia/dpac/consortium}). This study also made use of 
the SIMBAD database and the VizieR catalogue access tool, both operated at CDS, Strasbourg, France.


\software{  \texttt{Astropy} \citep{astropy2013,astropy2018}, 
            \texttt{SciPy} \citep{millman2011},
            \texttt{TOPCAT} \citep{taylor2005}, and 
            \textsc{StarGO} \citep{yuan2018}.
}
\clearpage
\bibliography{main}
\bibliographystyle{aasjournal}

\end{document}